%% file: main.tex
\documentclass[aps,prx,reprint,groupedaddress,nofootinbib]{revtex4-2}
\usepackage{float}
\input{macros}
\begin{document}

\title{Certified spectral functions from lattice Monte Carlo data}

\author{Sophie Mutzel}
\email[]{sophie.mutzel@minesparis.psl.eu}
\author{Antoine Tilloy}
\email{antoine.tilloy@minesparis.psl.eu}

\affiliation{Laboratoire de Physique de l’École Normale Supérieure, Mines Paris, Inria, CNRS, ENS-PSL, Centre Automatique et Systèmes (CAS),
Sorbonne Université, PSL Research University, Paris, France}

\date{\today}

\begin{abstract}
The Monte Carlo method, applied to lattice quantum field theory, gives access to Euclidean correlation functions with well-understood error bars. Recovering the observables one cares about, such as the spectral density, requires solving an ill-posed inverse problem, usually tackled with heuristics that lose rigorous control of the error. Instead of trying to find the ``best'' spectral density $\rho(\omega)$, we ask how small or large linear functionals $\int_{\mathbb{R}^+} G(\omega) \rho(\omega)  \upd \omega$ of it can be, given the Monte Carlo data and the reflection positivity of the lattice action. This is a convex but infinite-dimensional problem. We show how its dual can be rigorously relaxed into a hierarchy of finite semidefinite programs, solvable with standard solvers and enjoying strong convergence guarantees. The resulting bounds are rigorous even when the relaxation is not tight, and converge quickly to the regime where the error is entirely dominated by Monte Carlo statistics. The method also flags implausible Monte Carlo data, for instance underestimated error bars, through an infeasibility certificate. We demonstrate it on lattice $\phi^4$ theory in two dimensions.
\end{abstract}

\maketitle

\section{Introduction}
Monte-Carlo simulations have become the standard non-perturbative tool to solve strongly interacting field theories, from relativistic quantum field theory to condensed matter. In particular, lattice quantum chromodynamics (QCD) has achieved precision ab-initio computations of non-perturbative observables, including the light hadron spectrum~\cite{BMW:2008jgk}, light- and heavy-quark masses, decay constants and CKM matrix elements~\cite{FlavourLatticeAveragingGroupFLAG:2024oxs}, and a sub-percent determination of the leading hadronic vacuum polarization contribution to the muon anomalous magnetic moment~\cite{Boccaletti:2024guq}. 
These Monte-Carlo simulations provide systematically improvable access to Euclidean data, with statistical and systematic errors that can be quantified and reduced in a controlled manner.

The standard output of lattice simulations are Euclidean $n$-point correlation functions in imaginary time, from which physical observables, such as particle spectra, scattering amplitudes, transport coefficients, must be inferred. This requires analytical continuation back to Minkowski space, which, in practice, corresponds to inverting Laplace-type transforms, a notoriously ill-posed problem. Closely related inverse problems arise in condensed-matter physics, where dynamical response functions must be reconstructed from imaginary-time correlators. The simplest step towards real-time data, computing the spectral density from the $2$-point function, is already a difficult task:  small fluctuations in the input data can produce large variations in the reconstructed spectral density. 

Given the importance of this problem, a wide variety of techniques have been developed to address it. These include Bayesian and maximum-entropy approaches \cite{Jarrell:1996,Asakawa:2000tr,PhysRevLett.111.182003,beach2004identifyingmaximumentropymethod,Rothkopf:2022ctl,PhysRevD.95.056016}, Backus-Gilbert methods and extensions thereof \cite{BackusGilbert,Hansen:2017mnd,Hansen:2019idp}, Padé and rational approximations \cite{Tripolt:2018xeo,PhysRev.167.1411}, analytic continuation techniques \cite{Bergamaschi:2023xzx}, stochastic and neural-network based methods \cite{PhysRevB.57.10287,Buzzicotti:2023qdv,Wang:2021jou,PhysRevLett.124.056401,Horak:2021syv}, and more recently approaches based on Mellin transforms \cite{Bruno:2024fqc}. 

Most existing approaches aim to reconstruct a \emph{preferred} spectral density (under certain criteria) together with an estimate of its uncertainty. Although precise, the output of these methods does not enjoy the same properties as the initial ``raw'' correlation function data, which is unbiased and with well understood error bars. In the present work, we follow a different strategy, following Lawrence~\cite{Lawrence:2024hjm}: rather than attempting to determine a single ``best-fit'' spectrum, we characterize the entire set of spectral densities that are compatible with the Euclidean data and the positivity constraints implied by reflection positivity of the lattice action. This is an agnostic approach, that in principle allows to derive rigorous upper and lower bounds on arbitrary linear functionals of the spectral density, from the solution of a convex problem. 

The same philosophy of constraining observables via convex optimization built on physical principles underlies modern bootstrap programs: the conformal bootstrap \cite{Rattazzi:2008pe,Poland:2018epd} exploits unitarity and crossing of conformal field theory (CFT) four-point functions,  and the modern S-matrix bootstrap program \cite{Paulos:2016fap,Kruczenski:2022lot} exploits unitarity and analyticity of scattering amplitudes. More recently, this framework has been extended to incorporate ultraviolet data from Hamiltonian truncation through spectral densities \cite{Karateev:2019ymz,Chen:2021pgx}. 

The main novelty of our work is to rigorously relax the convex problem appearing in the spectral reconstruction problem into a hierarchy of finer and finer semi-definite programs (SDP), that can be solved efficiently with standard solvers, and with rigorous convergence certificates. The relaxation is tight in practice: with moderately fine levels of the hierarchy, the difference between the upper and lower bounds is completely dominated by Monte Carlo errors. 

Beyond the extraction of spectral information, the SDP framework we present also provides a non-trivial consistency test for Euclidean correlation functions themselves. If the problem is infeasible (the constraints cannot be satisfied simultaneously), as witnessed by rigorous certificates, the measured correlators are incompatible with the simultaneous assumptions of a positive spectral density and the error bars provided. This may reveal underestimated statistical uncertainties, residual auto-correlations, or other inconsistencies in the statistical treatment of Monte-Carlo data.

We first present more precisely our setup, then explain the convex programming idea and the sequence of steps that allow to turn it into a hierarchy of finite SDP, and then finally showcase the power of the method on real Monte-Carlo data for lattice $\phi^4$ theory in two dimensions.

\section{Setup}

\subsection{Monte Carlo data}

Our starting point may either be a Euclidean field theory discretized on a space-time lattice, or a condensed matter problem where space is natively discrete, and time is discretized. To fix the ideas, we consider the former setup here, in $d+1$ dimensions, with a lattice of size $L^{d-1}\times T$ for simplicity.

Such a model is specified by a lattice action $S$ which can be used to compute correlation functions. For example, for a scalar field one has:
\begin{equation}
  \langle\phi(\xb,t) \phi(0,0)\rangle  := \int\!\!\!\!\!\prod_{({\yb,\tau})\in \text{lattice}}\!\!\!\!\! \!\!\!\!\upd \phi(\yb,\tau) \; \phi(\xb,t) \phi(0,0)\,\e^{-\beta S(\phi)}\,.
  \label{eq:lattice_definition}
\end{equation}
Correlation functions are convenient to estimate in practice, for example with the Monte Carlo method. Assuming one has an efficient routine to sample fields exactly from the distribution $\exp(-\beta S(\phi))$, it is possible to approximate \eqref{eq:lattice_definition} by computing empirical averages over $N_\text{config}$ independent field configurations $\phi^{(r)}$:
\begin{equation}
  \overline{\phi(\xb,t) \phi(0,0)} := \frac{1}{N_\text{config}} \sum_{r=1}^{N_\text{config}} \phi^{(r)}(\xb,t) \phi^{(r)}(0,0) \, . 
  \label{eq:lattice_MC}
\end{equation}
We write $\langle \mathcal{O} \rangle$ the \emph{true} expectation values, and $\overline{\mathcal{O}}$ the empirical averages that are used as its estimators.

To reduce the empirical variance, it is standard to focus on the zero-momentum-projected two-point function
\begin{equation}\label{eq:2ptcorr_zero}
  C(t) \equiv \frac{1}{L^{d-1}} \sum_{\xb} 
\langle \phi(\xb,t)\,\phi(0,\vec{0}) \rangle,
\end{equation}
which can again be approximated by:
\begin{equation}\label{eq:2ptcorr_zero_empirical}
  \overline{C(t)} \equiv \frac{1}{L^{d-1}} \sum_{\xb} 
 \overline{\phi(\xb,t) \phi(0,\vec{0})},
\end{equation}
In the following, we assume that a reliable sampling method exists and that we thus have access to such empirical averages, that their fluctuations are approximately Gaussian, and that their covariance matrix is known. 

\subsection{Spectral density}
In the continuum, Euclidean two-point functions are related to the spectral density through a Laplace transform. An exact analogue holds on the lattice and is the starting point for our construction. 

Reflection positivity of the lattice action is equivalent to the existence of a positive Hermitian transfer matrix $\hat{\mathcal{T}} = e^{-a\hat{H}}$~\cite{Osterwalder:1974tc,montvay1994quantum} where $a$ is the lattice spacing and $\hat{H}$ some Hermitian Hamiltonian. This implies that the two-point function admits an exact Källén-Lehmann representation, even at finite spatial volume $L$, temporal extent $T$, and lattice spacing $a$,
\begin{equation}
  \label{eq:latticeKL}
  C(t) = \int_{\mathbb{R}^+}\mathrm{d}\omega\; \rho(\omega)\, K(\omega,t),
\qquad \rho(\omega) \geq 0,
\end{equation}
with a positive spectral density $\rho \equiv \rho_{\phi}$ that depends on the action only through the matrix elements of $\hat{\phi}$. Our framework applies to any observable $\mathcal{O}$ admitting such a positive spectral representation, but we focus on a scalar field on a lattice with periodic boundary conditions for simplicity.

The kernel in \eqref{eq:latticeKL} is then
\begin{equation}\label{eq:freeprop}
K(\omega,t) = e^{-t\omega} + e^{-(T-t)\omega},
\end{equation}
which reduces to the standard Källén-Lehmann representation (an inverse Laplace transform) in the $T \to \infty$ limit. By analogy with the continuum case we refer to $K(\omega,t)$ as the (zero-momentum projected) free propagator at temporal separation $t$.

The spectral density is a crucial object, \textit{e.g.} because it contains information about the particle masses and possible bound states of the model. Further, using equation \eqref{eq:latticeKL} and the expression of the propagator \eqref{eq:freeprop}, it can be used to extrapolate correlation functions to real-time. Our objective is now to assess how we can rigorously bound it from correlation functions with rigorous error bars, which is \emph{a priori} a difficult inverse problem.

\section{Rigorous bounds from convex programing} 
In this section we show how the inverse problem of extracting spectral information from Euclidean correlators can be formulated as a \emph{finite} semidefinite program, solvable with existing interior-point solvers.

\subsection{Primal formulation}
Since $\rho$ is a distribution rather than a function -- in finite volume it is a sum of Dirac delta functions -- it is hopeless to try to bound it pointwise. Rather, we should consider smeared linear functionals of the form
\begin{equation}\label{eq:smeared_rho}
  (G,\rho):= \int_{\mathbb{R}^+} G(\omega)\,\rho(\omega)\,\upd \omega,
\end{equation}
where $G$ is a test function. The simplest choice is a rectangular window, $G(\omega) = \mathbf{1}_{[\omega_1,\omega_2]}(\omega),$
in which case the functional reduces to the spectral density integrated over the interval $[\omega_1,\omega_2]$. More generally, we could consider a Gaussian $g_{\omega_0}^\sigma$ of width $\sigma$ centered around $\omega_0$, or, for reasons that will be made clearer later, any function that is easily bounded by piecewise affine functions.

Intuitively, the upper and lower bounds on $(G,\rho)$ are obtained by maximizing and minimizing over the admissible set of spectral densities, under the constraint that the corresponding correlation functions are compatible with the data.

Reflection positivity of the lattice action implies that the admissible set of spectral density is simply positive distributions: 
\begin{equation}
  \forall \omega \in \mathbb{R}^+, \;\, \rho(\omega)\geq 0\;.
\end{equation}

Encoding the statistical compatibility between the correlation functions obtained from the lattice Källén-Lehmann representation \eqref{eq:latticeKL} and the Euclidean Monte-Carlo data is slightly more subtle. To this end, we introduce the residuals $y_j$ associated to each of the $N_c = T/2+1$ distinct possible (imaginary) time $t_j$ separations: 
\begin{equation}
  y_j = \int_{\mathbb{R}^+} K(\omega , t_j)\,\rho(\omega)\,\mathrm{d}\omega - \bar{C}(t_j)\, .
  \label{eq:residuals}
\end{equation}
Assuming the Monte Carlo estimates are approximately Gaussian-distributed around their true means with covariance $\Sigma$, the residual $y_j$ is itself approximately Gaussian, and the squared Mahalanobis distance $y^T \Sigma^{-1} y$ follows a $\chi^2_{T/2+1}$ distribution. Requiring \begin{equation}\label{eq:Fmaxconstraint}
y^T \Sigma^{-1} y \leq F_{\max}
\end{equation}
restricts $\rho$ to those spectral densities statistically compatible with the data at a chosen confidence level.

To find a meaningful threshold $F_\text{max}$ we follow Lawrence~\cite{Lawrence:2024hjm} and build the empirical distribution of the squared Mahalanobis distance from the data. 
 From the $N_{\rm config}$ individual field configurations we generate $N_\text{sample}$ bootstrap (or jackknife) resamples. For each resample $k$, we compute the connected correlator $\braket C^{(k)}(t_j)$ where $\braket{\cdot}$ denotes the average over the resampled configurations. The covariance matrix $\Sigma$ is estimated from the spread of these bootstrap correlators, and the squared Mahalanobis distance
  \begin{equation}\label{eq:Fk}
    F_k = \left(\mathcal{C}^{(k)} - \bar{\mathcal{C}}\right)^T \Sigma^{-1} \left(\mathcal{C}^{(k)} - \bar{\mathcal{C}}\right)
  \end{equation}
is evaluated for each resample, where $\cC$ is a compact notation for the vector $\{\braket{C}(t_j)\}_j$, the index $k = 1,\ldots,N_\text{sample}$ labels the sample and $\bar{\mathcal{C}}$ is the ensemble mean. It is then natural to define $F_{\max}$ as the $99\%$ quantile of this empirical cumulative distribution. Since $F_{\max}$ is constructed purely from the Monte Carlo data, it is independent of the inversion problem and can be computed prior to the optimization.

Putting the two together gives the convex problem 
\begin{subequations}\label{eq:primal}
\begin{align}
 p^*= \min \limits_{\rho} \quad & \int_{\mathbb{R}^+} G(\omega) \; \rho(\omega) \;\d \omega \label{eq:primal_cost}\\ 
\text { under } 
& \cdot \rho(\omega) \geq 0 \;,\qquad \forall \omega \geq 0\label{eq:primal_pos}\\  
& \cdot  y(\rho)^T \Sigma^{-1} y(\rho) \leq F_{\max} \label{eq:Fmaxconstr} \, ,
\end{align}
\end{subequations}
which gives a lower bound for $(G,\rho)$ (the same problem with a $\max$ giving an upper bound instead).

To obtain this convex optimization problem, we have closely followed the steps of Lawrence ~\cite{Lawrence:2024hjm}. We now differ in the way we treat the constraint \eqref{eq:Fmaxconstr}. In~\cite{Lawrence:2024hjm}, it was kept as a non-linear constraint, which then gave a dual problem with non-linear objective. Lawrence solved it with an \emph{ad hoc} solver, which, though numerically efficient, does not provide rigorous certificates. Instead, we aim to obtain a finite semi-definite program, which comes with rigorous guarantees of convergence, efficient implementations in standard packages, and an ability to extract rigorous certificates for the bounds.

To this end, we use the Schur complement trick and introduce the $(N_c+1) \times (N_c+1)$ matrix
\begin{equation}\label{eq:PSDmatrixA}
A(y)=\begin{aligned}
\left(\begin{array}{ll}
F_{\max} & y^T \\
y & \Sigma 
\end{array}\right) \, .
\end{aligned}
\end{equation}
The inequality \eqref{eq:Fmaxconstr} is equivalent to the positive-semidefiniteness (PSD) of $A$:
\begin{equation}
y^T\Sigma^{-1}y\leq F_{\max} \quad \Leftrightarrow \quad A\succeq 0 \;.
\end{equation}
Since $\rho \mapsto A(y(\rho))$ is an affine function of $\rho$, the problem \eqref{eq:primal} with \eqref{eq:Fmaxconstr} replaced by $A(y(\rho))\succeq 0$:
\begin{subequations}\label{eq:primal_postSchur}
\begin{align}
 p^*= \min \limits_{\rho} \quad & \int_{\mathbb{R}^+} G(\omega) \; \rho(\omega) \;\d \omega \label{eq:primal_cost_postschur}\\ 
\text { under } 
& \cdot \rho(\omega) \geq 0 \;,\qquad \forall \omega \geq 0\label{eq:primal_pos_postschur}\\  
& \cdot  A(y(\rho))\succeq 0 \label{eq:Apos} \, ,
\end{align}
\end{subequations}
 is a semidefinite program.

\begin{figure}
    \centering
    \includegraphics[width=1\linewidth]{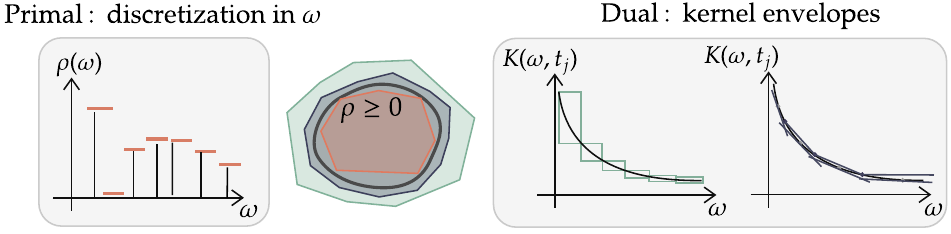}
    \caption{\emph{Middle}: Schematic projection of the convex space of admissible positive spectral densities. 
\emph{Left}: Approximation by discretizing in $\omega$, which is fine for all practical purposes, if a large number of variables $N_v$ for discretization in $\omega$ is taken. The resulting bounds approach the true ones from inside as the number of grid points $N_v$ grows, but are not rigorous at finite $N_v$ (orange). 
\emph{Right}: Rigorously bounding the Källén-Lehmann kernel by 
piecewise constant (turquoise) or piecewise affine (dark blue) envelopes enlarges the feasible set. The resulting bounds approach the true ones from outside as the number of envelope pieces grows, and are rigorous at every finite $N_v$.}
    \label{fig:convex}
\end{figure}

This ``\emph{primal}'' SDP is unfortunately still infinite-dimensional: the optimization variable is a continuous function $\rho(\omega) \geq 0$.  A standard approach would be to discretize $\rho$ on a grid $\{\omega_n\}_{n=1}^{N_v}$ and solve the resulting finite SDP with a standard solver. However, the bounds obtained this way are only rigorous up to a discretization error (left of Fig.~\ref{fig:convex}). To obtain rigorous bounds, we move to a dual formulation which is then easier to rigorously \emph{relax} into a finite SDP.

\subsection{Dual formulation}
The primal formulation identifies the extremal spectral densities compatible with the Monte Carlo data. By construction, any feasible primal point corresponds to a candidate spectral density and therefore provides an upper bound to the minimum (or lower bound to the maximum) of the optimization problem. Hence, to get a useful bound, one needs to \emph{attain} the optimum exactly. It would be more convenient to have a dual formulation, where every feasible point provides a lower bound to the minimum (or an upper bound to the maximum), thus providing useful bounds on $(G,\rho)$ even without perfect optimization.

One can obtain a dual problem to \eqref{eq:primal} using Langrange duality: introducing Lagrange multipliers for the positivity and consistency constraints, and optimizing over the spectral density to get a problem where the multipliers are the remaining variables. We show how to obtain the dual in this principled way in App.~\ref{app:dualfromlagrangian}. Here, we show that it can be derived directly, in a rather intuitive way.

To this end, it is convenient to decompose the $\rho$-independent part of the PSD constraint \eqref{eq:PSDmatrixA} into a constant contribution and an $\omega$-dependent contribution,
\begin{equation}
\begin{aligned}
&\cA _c =\left(\begin{array}{cc}
F_{\max } & -\bar{\mathcal{C}}^T \\
-\bar{\mathcal{C}} & \Sigma
\end{array}\right)\;,
\end{aligned}\; \begin{aligned}
&\cA (\omega) =\left(\begin{array}{cc}
    0 & \mathcal{K}(\omega)^T \\
    \mathcal{K}(\omega) & [0]_{N_c\times N_c}
\end{array}\right)\;,
\end{aligned}\nonumber
\end{equation}
where the vectors $\bar{\mathcal{C}}=\{ \bar{C}(t_j)\}_{j=1}^{N_c}$ and $\mathcal{K}(\omega) = \{ K(\omega,t_j)\}_{j=1}^{N_c}$, yielding
\begin{equation}
A(y(\rho))=\cA_c+\int_{\mathbb{R}^+} \cA(\omega) \rho(\omega) \;\d \omega \;.
\end{equation}
Now, suppose there exists a $(N_c+1)\times(N_c+1)$ dimensional PSD matrix $\Gamma$ such that
\begin{equation}\label{eq:sos1_bis}
G(\omega)  - \tr \l[\Gamma  \cA(\omega)  \r] \geq 0 \qquad \forall \omega \in \mathbb{R}^+ \;.
\end{equation}
Integrating against the ``true'' non-negative spectral density $\rho$ implies 
\begin{equation}
\int_{\mathbb{R}^+} G(\omega) \rho(\omega)\, \d \omega \geq \int_{\mathbb{R}^+} \tr[\Gamma \cA(\omega)]\rho(\omega)\, \d \omega \;.
\end{equation}
Using $A(y)\succeq0$ and $\Gamma\succeq0$, we furthermore have $\tr[\Gamma A(y)]\geq 0$, which implies
\begin{equation}\label{eq:sos3}
  \int_{\mathbb{R}^+} \tr[\Gamma \cA(\omega)]\rho(\omega)\, \d \omega \geq -\tr[\Gamma\cA_c] \; .
\end{equation}
Combining the two inequalities \eqref{eq:sos1_bis} and \eqref{eq:sos3} yields the certificate
\begin{equation}
  \int_{\mathbb{R}^+} G(\omega)\rho(\omega)\, \d \omega \geq -\tr[\Gamma\cA_c] \;.
\label{eq:sos1}
\end{equation}
Hence, going back to our original assumption, every matrix $\Gamma$ satisfying Eq.~\eqref{eq:sos1_bis} provides a rigorous lower bound on the smeared spectral density. The strongest such bound is obtained by maximizing the right-hand side of Eq.~\eqref{eq:sos1}, leading to the following dual optimization problem 
\begin{subequations}\label{eq:dualSDP}
\begin{align}
d^*=\max\limits_{\Gamma} \quad &  - \tr \l[\Gamma \cA_c \r] \label{eq:costdual}\\ 
\text { under } 
& \cdot \Gamma \succeq 0\\
& \cdot G(\omega)  - \tr \l[\Gamma  \cA(\omega)  \r] \geq 0  \quad \; \forall \omega \in \mathbb{R}^+  \label{eq:positivitydual}
\end{align}
\end{subequations}
which is indeed the dual of \eqref{eq:primal} in the sense of convex optimization. Any feasible point of Eq.~\eqref{eq:dualSDP} therefore immediately yields a certified lower bound. In the present setting strong duality holds and the optimum of the dual problem coincides with that of the primal problem,
\begin{equation}
d^\ast = p^\ast \;.
\end{equation}
This dual formulation gives us certificates, but it still cannot be solved easily because the positivity constraint \eqref{eq:positivitydual} has to hold for all $\omega\in\mathbb{R}^+$. Discretizing naively this constraint is not better than discretizing the primal variables. Instead, we need to find a relaxation of this problem, \textit{i.e.} a finite set of constraints that imply positivity for all $\omega\geq0$. We do so by carefully bounding the propagator $K$.

\subsection{Rigorous relaxation}\label{sec:relaxation}
The dual formulation \eqref{eq:dualSDP} provides rigorous bounds whenever the positivity condition \eqref{eq:positivitydual} can be verified exactly for all $\omega \in \mathbb{R}^+$. We now show how to enforce this condition \emph{rigorously} on a finite set of constraints, by bounding the kernel $K(\omega,t_j)$ in the Källén-Lehmann relation by upper and lower 
bounding functions $f_j^\pm(\omega)$. 
The resulting problem is a finite-dimensional SDP whose feasible set strictly contains that of the original problem (a genuine relaxation) so that any feasible dual point still certifies a rigorous lower bound on the primal objective.

We now assume that we can find functions $f_j^\pm(\omega)$ bounding the propagator from above and below\footnote{Some kernels $\omega \mapsto K(\omega,t_j)$ diverge as $\omega \to 0$. On a finite lattice the spectrum of the transfer matrix is discrete and bounded below by the lowest excitation energy $\omega^\ast > 0$ above the vacuum, provided one works with connected correlators and no exact zero modes are present. 
In the large-volume regime $\omega^\ast$ approaches the lightest physical mass. 
For the kernel \eqref{eq:freeprop} considered below, $K(\omega,t_j)$ remains finite at $\omega = 0$, and rigorous bounds can be constructed on the entire interval $\omega \geq 0$.}
\begin{equation}\label{eq:envelopes}
f_j^-(\omega) \leq K(\omega,t_j) \leq f_j^+(\omega), \qquad \forall \omega > 0,
\end{equation}
for every $j = 1,\ldots,N_c$. We further assume that it is straightforward to verify the positivity for all $\omega\geq 0$ of linear combinations of such functions (we will later construct them explicitly as piecewise constant or piecewise affine functions). 

The idea is now to see the residuals $y_j$ not as functions of $\rho$ but as proper variables. The primal problem \eqref{eq:primal_postSchur} is indeed equivalent to 
\begin{subequations}\label{eq:primal_slacked}
\begin{align}
 p^*= \min \limits_{\rho,y} \quad & \int_{\mathbb{R}^+} G(\omega) \; \rho(\omega) \;\d \omega \label{eq:primal_cost_slacked}\\ 
\text { under } 
                                  & \cdot \rho(\omega) \geq 0 \;,\qquad \forall \omega \geq 0\label{eq:primal_pos_slacked}\\  
& \cdot  A(y)\succeq 0 \label{eq:Apos_slacked} \, ,\\
&\cdot  y_j =\int K(\omega,t_j)\rho(\omega) \upd\omega - \bar{C}(t_j)  \label{eq:equality_slack} 
\end{align}
\end{subequations}
Using \eqref{eq:envelopes}, we can relax the last equality \eqref{eq:equality_slack} into a pair of inequalities:
\begin{subequations}\label{eq:primal_relaxed}
\begin{align}
  p^*_\text{rel}= \min \limits_{\rho,y} \quad & \int_{\mathbb{R}^+} G(\omega) \; \rho(\omega) \;\d \omega \label{eq:primal_cost_relaxed}\\ 
\text { under } 
                                  & \cdot \rho(\omega) \geq 0 \;,\qquad \forall \omega \geq 0\label{eq:primal_pos_relaxed}\\  
& \cdot  A(y)\succeq 0 \label{eq:Apos_slacked_relaxed} \, ,\\
&\cdot  y_j \leq\int f_j^+(\omega) \rho(\omega) \upd\omega - \bar{C}(t_j)  \label{eq:inequality_relaxed_plus} \\ 
&\cdot  y_j \geq\int f_j^-(\omega) \rho(\omega) \upd\omega - \bar{C}(t_j)  \label{eq:inequality_relaxed_minus} 
\end{align}
\end{subequations}
Since the constraints are now weaker, the feasible set is larger, and thus $p^{*}_\text{reg} \leq p^{*}$: it is a genuine relaxation, and we keep rigorous bounds.

This relaxed primal \eqref{eq:primal_relaxed} also admits a relaxed dual, which can be found through Lagrange duality
\begin{subequations}\label{eq:dualrelaxed}
\begin{align}
  d^*_\text{rel}=\max\limits_{\Gamma,\; \alpha} \quad & - \tr \l[ \Gamma \cA_c\r]  \label{eq:cost}\\ 
\text { under } 
                                                      & \cdot G(\omega) +\sum_j \alpha_j f^-_{j}(\omega) \label{eq:relaxeddual_positivity} \\
& -\sum_j (\alpha_j+2 \gamma_{0,j}) f^+_{j}(\omega)  \geq 0 \;, \quad \forall \omega \in \mathbb{R}^+ \nonumber\\
& \cdot \alpha_j + 2 \gamma_{0,j} \geq 0 \;, \quad j=1,\ldots,N_c \label{eq:positivityrelaxed}\\
& \cdot  \alpha_j \geq 0  \;, \quad j=1,\ldots,N_c \label{eq:relationgammaalpha}\\
& \cdot  \Gamma \succeq 0 \;, \label{eq:GammaSDP}
\end{align}
\end{subequations}
where we have pararameterized $\Gamma$:
\begin{equation}\label{eq:Gamma}
\Gamma = \begin{pmatrix} \gamma_{00} & \gamma_0^T \\ 
\gamma_0 & \Gamma_\Sigma \end{pmatrix}.
\end{equation}
The intuitive justification is analogous to that of the previous section. For any $\Gamma \succeq 0$ and $\alpha_j \geq 0$ satisfying \eqref{eq:positivityrelaxed} and the auxiliary positivity conditions \eqref{eq:relationgammaalpha}, multiplying by $\rho(\omega) \geq 0$ and integrating yields
\begin{equation}
\int G(\omega)\,\rho(\omega)\,\mathrm{d}s \geq -\mathrm{tr}\bigl[\Gamma\,\mathcal{A}_c\bigr],
\end{equation}
where the negative-sign contributions from $f^+$ are absorbed by the $\alpha_j + 2\gamma_{0,j} \geq 0$ multiplier, exactly cancelling the $y_j$-coupling generated by $\Gamma$ acting on the off-diagonal block of $A(y)$. Every feasible point of \eqref{eq:dualrelaxed} therefore certifies a lower bound on the smeared spectral density. Now all that remains is to turn this relaxed dual into a finite SDP, by choosing a basis of functions where \eqref{eq:relaxeddual_positivity} becomes a finite set of linear inequalities.

\subsection{Explicit envelope constructions}

We now show explicit examples of envelope functions $f_j^{\pm}$ that tightly bound the propagator. The only assumption we need is that $\omega \mapsto K(\omega,t_j)$ is \emph{convex}. This is satisfied by the lattice kernel \eqref{eq:freeprop} (a sum of exponentials) and, more generally, appears to hold for the kernels arising in (lattice or continuum) Källén-Lehmann representations\footnote{For kernels that are not convex, one can often restore convexity by a redefinition of $\rho$, to absorb a positive but non-convex factor. Otherwise, our method applies to kernels that are piecewise convex.}. 

\begin{figure}
\centering
\includegraphics[width=\linewidth]{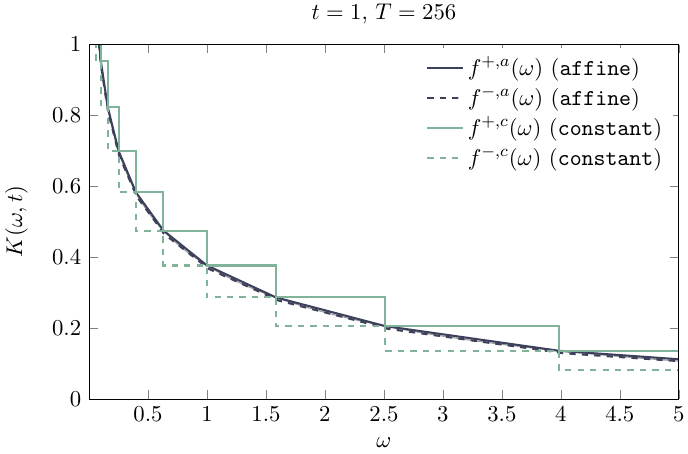}
\caption{Rigorous piecewise constant (turquoise) and piecewise affine (dark blue) upper and lower envelopes $f_j^\pm(\omega)$ for the lattice kernel \eqref{eq:freeprop} of the Källén-Lehmann relation at one representative $t_j$. 
Convexity of $K(\omega,t_j)$ on $(0,\infty)$ ensures that secants provide rigorous upper and tangents provide rigorous lower bounds.}
\label{fig:bounds}
\end{figure}

First, a function that is monotonous, like the propagator we consider, can be bounded by piecewise constant functions. On an interval $[\omega_1,\omega_2]$ the values taken by the function at the boundaries provide upper and lower bounds to the values everywhere. Better, for a convex function, we can take the secant as upper bound, and any tangent (for example, the tangent at the midpoint) as lower bound\footnote{Note that with this construction, the upper bound is a continuous function, while the lower bound is not. This does not matter in practice.}. This latter option provides far tighter bounds as we illustrate in Fig. \ref{fig:bounds}, but we discuss the piecewise constant case as well as it is conceptually easier.

To make the construction explicit, we take $N_v-1$ points $0< \omega_1 < \cdots < \omega_{N_v-1}$ logarithmically spaced in $\mathbb{R}^+$, and associate two sets of functions
\begin{align}
  \texttt{constant} \;\;\; & b^{\rm{c}}_k(\omega) = \mathbf{1}_{[\omega_{k-1},\,\omega_{k}]}(\omega) \; \\ 
  \texttt{affine} \;\; & \left\lbrace\begin{array}{l}
      b_k^{\rm a,1}(\omega) = \mathbf{1}_{[\omega_{k-1}, \omega_{k}]}(\omega)\\
      b_k^{\rm a,2}(\omega) = (\omega - \omega_{k-1})\,\mathbf{1}_{[\omega_{k-1}, \omega_{k}]}(\omega),
  \end{array}\right.
\end{align}
with the convention $\omega_0=0$ and $\omega_{N_v} = +\infty$. Using the construction above, we can thus find explicit bounding envelopes written as linear combinations of functions in this set:
\begin{align}\label{eq:basisexpansion}
  f_j^{\pm,\mathrm{c}}(\omega) &= \sum_{k=1}^{N_v} f_{j,k}^\pm\, b^c_k(\omega), \\ 
  f_j^{\pm,\mathrm{a}}(\omega) &= \sum_{k=1}^{N_v} f^{1,\pm}_{j,k}\, b^{\rm a,1}_k(\omega) + f^{2,\pm}_{j,k}\, b^{\rm a,2}_k(\omega) \, .
\end{align}
In both cases, the numerical coefficients in this expansion are precomputed from the propagator (we detail how in App. \ref{app:finitedimrelaxed}), and are fixed during the optimization.

Finally, we assume that the test function $G$ we care about is \emph{exactly} piecewise constant or piecewise affine (depending on which basis is chosen), or at least easily tightly bounded by functions $G^\pm$ that are of this form. For simplicity, we assume the former, and have coefficients $G_k$ or $G_k^{1}, G_k^{2}$ depending on the basis chosen.

It is now clear that the pointwise constraint~\eqref{eq:positivityrelaxed} reduces to a finite number of linear inequalities. In the piecewise constant case \eqref{eq:positivityrelaxed} is equivalent to
\begin{equation}\label{eq:inequalities_constant}
  \forall k,\quad G_k + \sum_j \alpha_j f^{-}_{j,k} - (\alpha_j +2\gamma_{0,j}) f^{+}_{j,k} \geq 0 \, ,
\end{equation}
and in the piecewise affine case it is equivalent to
\begin{align}\label{eq:inequalities_affine}
  \forall k,&\quad G_k^{1} + \sum_j \alpha_j f^{1,-}_{j,k} - (\alpha_j +2\gamma_{0,j}) f^{1, +}_{j,k} \geq 0 \\
\begin{split} 
            &G_k^{1} + (\omega_{k} - \omega_{k-1}) G_k^2+ \sum_j \alpha_j f^{1,-}_{j,k} - (\alpha_j +2\gamma_{0,j}) f^{1, +}_{j,k}    \\ 
            +&(\omega_{k} - \omega_{k-1}) \Big[\sum_j \alpha_j f^{2,-}_{j,k} - (\alpha_j +2\gamma_{0,j}) f^{2, +}_{j,k}\Big] \geq 0 \, .
\end{split}\nonumber
\end{align}
In both cases, these are a finite set of linear inequalities. The relaxed dual \eqref{eq:dualrelaxed} is therefore a genuinely finite-dimensional SDP, solvable by standard interior-point methods, and every feasible solution it returns (optimal or not) is a rigorous lower bound (see right of Fig.~\ref{fig:convex}).

The relaxation gap closes as our discretization gets finer ($N_v \to \infty$) and we will see in the next section that already moderate $N_v$ (of a few hundreds) produces tight bounds for realistic lattice data, at least with piecewise affine envelopes.

\section{Application}
We now apply the method to real lattice Monte-Carlo data. First, we demonstrate that the SDP formulation can be implemented in a realistic lattice field theory setting, and second, we study the convergence of the resulting bounds as a function of both the relaxation and the statistical precision of the input data. 

\subsection{Lattice field model}
As a benchmark, we consider two-dimensional $\phi^4$ theory. Although among the simplest interacting quantum field theories, it has a non-trivial spectral density while at the same time remaining sufficiently well understood to allow a study of the resulting bounds. 

Specifically, we study $\phi^4$ theory in two Euclidean spacetime dimensions, with continuum action
\begin{equation}
S_E[\phi]=\int_{\mathbb{R}^2}\!\mathrm{d}^2x\;\Big[\tfrac{1}{2}(\nabla\phi)^2+\tfrac{m_0^2}{2}\,\phi^2+g\,\phi^4\Big],
\end{equation}
which we discretize on a square lattice of spacing $a$ in the standard way \cite{Delcamp:2020hzo,Loinaz:1997az,Schaich:2009,Bosetti:2015,Bronzin:2019},
\begin{equation}
S_E^a[\phi]=\sum_{\langle i,j\rangle}\frac{(\phi_i-\phi_j)^2}{2}
+\sum_i\Big[\frac{m_b^2}{2}\,\phi_i^2+\lambda\,\phi_i^4\Big],
\end{equation}
with lattice spacing $a$ and where $\langle i,j\rangle$ runs over nearest-neighbour pairs and the dimensionless bare parameters are $m_b=a\,m_0$ and $\lambda=a^2 g$. 

In two dimensions, the theory is super-renormalizable: by power counting the only superficially divergent one-particle-irreducible amplitude is the one-loop self-energy (the tadpole diagram), which is logarithmically divergent. A single mass counterterm therefore suffices to renormalize the theory. We fix the bare mass through the one-loop tadpole subtraction
\begin{equation}\label{eq:massren}
m_b^2:=m_{\rm ren}^2-12\,\lambda\,A(m_{\rm ren}^2),
\end{equation}
where $m_{\rm ren}$ is the renormalized (normal-ordered) lattice mass and $A(m_{\rm ren}^2)$ is the lattice tadpole,
\begin{equation}
\begin{split}
 A(m_{\rm ren}^2)&=\frac{1}{LT}\sum_{k_1=0}^{T-1}\sum_{k_2=0}^{L-1}\\
&\quad \frac{1}{m_{\rm ren}^2+4\big[\sin^2(\pi k_1/T)+\sin^2(\pi k_2/L)\big]} \;.
\end{split}
\end{equation}
The finite part of this subtraction is a scheme choice. Subtracting the full lattice tadpole $A(m_{\rm ren}^2)$ is the special one that coincides with normal ordering the continuum theory, so that the renormalized parameters $m_{\rm ren}$ and $\lambda$ directly define the continuum normal-ordered $\phi^4$ model and its physical quantities in the limit $a\to 0$ \cite{Loinaz:1997az}. This model is characterised by the single dimensionless coupling
$f\equiv g/m^2=\lambda/m_{\rm ren}^2$, and since $\lambda=a^2 g$, $m_{\rm ren}=a\,m$ at fixed physical $f$, the continuum limit is reached along a line of constant $f$ with $a\to 0$ or equivalently $\lambda\to0$.

We simulate this theory on a $L\times T =256\times256$ periodic lattice with bare parameters fixed by Eq.~\eqref{eq:massren}.  Our Monte Carlo method is standard, and we do not introduce any particular innovation. The update algorithm consists of one Gaussian heatbath sweep, four over-relaxation sweeps, and a global $\mathbb{Z}_2$ sign flip $\phi(x)\to-\phi(x)$ which together form one update. Consecutive measurements are separated by 40 such updates; on each of these configurations we measure the connected zero-momentum projected two-point correlator $C(t)$, Eq.~\eqref{eq:2ptcorr_zero}. We verify that the integrated autocorrelation time of the zero-momentum correlator is small compared to this separation. 
From those measurements we obtain $\bar{C}(t_j)$, the mean value of the correlator for each time slice $t_j$ (symmetrized in time), and the statistics $F_{\max}$.

\subsection{Convergence results}
To obtain the bounds discussed below, we wrote the dual SDP \eqref{eq:dualrelaxed} (with the constraints \eqref{eq:inequalities_affine}) in \texttt{Julia}, using the framework \texttt{JuMP} \cite{Lubin2023}. We then used \texttt{Mosek} \cite{mosek}, a primal-dual interior point solver, to solve the resulting optimization problem (we checked that \texttt{Hypatia} \cite{coey2021solvingnaturalconicformulations}, another interior point solver, provided identical results). For the field theory we consider, and at the level of precision reachable with our Monte Carlo data, solving the SDP in double precision seemed sufficient.

\begin{figure}
    \centering
    \includegraphics[width=1\linewidth]{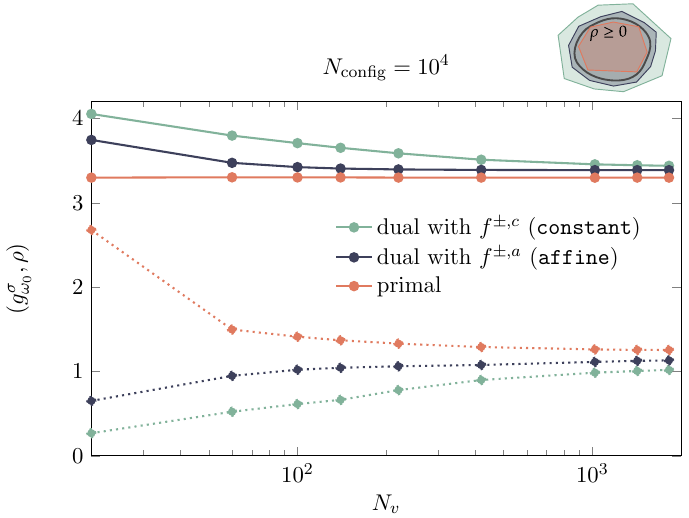}
    \caption{
    Convergence with the number of \texttt{constant} or \texttt{affine} basis functions $N_v$ for the envelopes $f^\pm$ of the kernel $K(\omega,t)$. 
    We used $10^4$ configurations for $\phi^4$-theory at $\lambda = 4 \cdot 10^{-3}$ and $g = 2$, on a $256 \times 256$ lattice. We also show the bounds obtained by solving the primal SDP \eqref{eq:primal} discretized on a grid with $N_v$ discretization points.
    } 
\label{fig:comparison_boundings} 
\end{figure}

We first study the convergence as a function of the coarseness of the relaxation. We pick one strongly coupled point of the $\phi^4$ model at $\lambda = 4 \cdot 10^{-3}$ (corresponding to a lattice spacing $a \approx 0.045$) and bare mass $m_\text{b}^2 = 2 \cdot 10^{-3}$. In the continuum and infinite-volume limit ($a \to 0$, $L,T \to \infty$) these parameters correspond to the normal-ordered continuum theory at $g \simeq 2$. As observable, we choose $(g^\sigma_{\omega_0}, \rho)$, where $g^\sigma_{\omega_0}$ is a piecewise constant approximation of a Gaussian of width $\sigma = 0.1$ and centred at $\omega = \sqrt{0.2}$ (which is approximately where we expect the mass peak). We display in Fig.~\ref{fig:comparison_boundings} the rigorous dual bounds as a function of the number of basis functions $N_v$, for both piecewise constant and piecewise affine envelopes. For comparison, we also include the results obtained from a direct discretization of $\rho$ in the primal formulation, with the same number $N_v$ of variables. The corresponding bounds are not rigorous bounds for the original problems, but they are respectively lower bounds to the dual upper bound, and upper bounds to the dual lower bound. They thus give a good idea of convergence. Note however that we expect our dual bounds with affine functions to converge much faster as they are of a higher order than naive discretization.

This is indeed what we observe: the piecewise affine envelopes (secants and tangents) are  tighter than the piecewise constant ones at fixed $N_v$, the primal discretized bounds approach the rigorous dual bounds from the inside. 
The piecewise affine envelopes converge extremely quickly: already at $N_v \sim$ a few hundred basis functions the bounds stabilize, and further refinement of the envelopes leaves them unchanged,
which suggests that the relaxation gap closes fast in practice. Note again, that even when they are not converged as a function of $N_v$, the dual bounds are rigorous.

\begin{figure}
    \centering
    \includegraphics[width=1\linewidth]{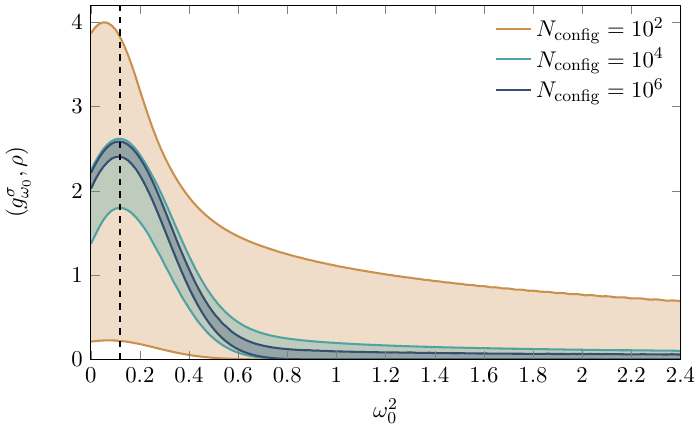}
    \caption{
    Rigorous upper and lower bounds on the smeared spectral density ($\sigma^2 = 0.2$) from the dual SDP~\eqref{eq:dualrelaxed} for $\phi^4$-theory at $\lambda = 4 \cdot 10^{-3}$ and $g = 2$, on a $256 \times 256$ lattice. 
    The symmetrized correlator~\eqref{eq:2ptcorr_zero_empirical} provides $N_c = 129$ constraints. The vertical dashed line corresponds to the mass-gap estimate, obtained from an effective-mass fit.}
\label{fig:reconstruction_Nconfig} 
\end{figure}

We also verify the convergence as a function of the number of Monte Carlo configurations, with a relaxation sufficiently fine to have negligible effects (dual with piecewise affine basis functions, and $N_v=700$ subdivisions). The results are displayed in Fig.~\ref{fig:reconstruction_Nconfig} and show the rigorous upper and lower bounds on the smeared spectral density $(g^\sigma_{\omega_0}, \rho)$ as a function of $\omega_0^2$, for $\sigma =  \sqrt{0.2}$, and for the same parameters of the model. 

As expected, the bounds get tighter as the number of configurations $N_{\rm config} \in \{10^2, 10^4, 10^6\}$ is increased. At $N_{\rm config} = 10^4$, the bounds are already tight enough to resolve the single-particle peak well. The position of the peak agrees with the mass-gap estimate from an effective-mass fit shown for comparison. However, while the method allows to bound the ``continuum'' part of the spectrum from above, our Monte Carlo data is insufficiently precise to bound it from below.

\subsection{Inconsistency certificates}
It may happen that the constraints in \eqref{eq:primal} cannot be satisfied simultaneously, or equivalently, that the dual \eqref{eq:dualrelaxed} is unbounded (the bounds cross). This is the sign that no spectral density is compatible with the Monte Carlo data. For a reflection positive action, if a given $(\bar{C}, \Sigma)$ is inconsistent with the existence of \emph{any} positive spectral density at the $99\%$ level, the underlying statistical assumptions are falsified. Our procedure therefore not only produces bounds on integrals of $\rho$, but also certifies whether a specific error treatment of the Monte Carlo data is itself consistent with the most basic physical constraint. 

To illustrate this fact, we generated intentionally flawed Monte Carlo data by violating standard simulation practice. We ran 100 short Markov chains of 100 measurements each, starting from random (unthermalized) field configurations and measuring after every single (full) update without any de-correlation gap. The resulting $10^4$ measurements were then combined and analyzed as if they were independent samples, leading to severely underestimated statistical errors: the early configurations are far from equilibrium, and consecutive measurements exhibit strong autocorrelations that are not accounted for in the naive error estimate. Following now standard practice~\cite{Lawrence:2024hjm}, we further regularized the empirical covariance matrix according to 
\begin{equation}
\Sigma \rightarrow \Sigma_{\rm reg}
=
\Sigma+\epsilon \lVert \Sigma \rVert \mathbf{1}\;,
\end{equation}
with $\epsilon=10^{-6}$ (see App. \ref{app:regularization} for more detail). 

With such a flawed Monte Carlo simulation, we find that the resulting SDP is infeasible. In other words, no positive spectral density exists that is compatible with this measured correlator \emph{within the quoted uncertainties}.  While this example is intentionally extreme, it illustrates an important feature of the method. The SDP does not merely produce bounds on spectral observables, it also provides a non-trivial consistency test of the Monte-Carlo data and its estimated covariance matrix.

\section{Discussion}
In this article,  following Lawrence~\cite{Lawrence:2024hjm}, we started by showing that bounding spectral densities with Monte Carlo data can be expressed as a convex optimization problem. We then moved from this convex problem to an infinite dimensional SDP using a Schur complement trick. To get rigorous certificates, even if the optimization has not converged, we showed that one could easily move to a (still infinite) dual formulation. Finally, we relaxed this dual problem by carefully bounding the propagator with functions belonging to a simple basis, which provided a finite SDP rigorously bounding the solutions of the original convex problem we started with. This is the main advantage of our approach: provided the Monte Carlo data is correct, and its error well estimated, our bounds are rigorous, even if the relaxation is not tight. In practice, we observed that, at least for reasonable Monte Carlo data, our relaxation is indeed tight, meaning that the errors are dominated by the Monte Carlo error bars. Hence, on top of being rigorous, our bounds are the best one can get with the minimal hypothesis we made. 

Our approach also allows us to rigorously show that some Monte Carlo data is incorrect, in the sense that its data and associated error bars are incompatible with the simulation of a certain reflection positive field theory. Since positivity of the spectral density and covariance-based error estimates underlie many lattice analyses, such consistency checks are valuable independently of the spectral reconstruction problem itself. Admittedly, we do not yet have a very stringent test, and one needs quite some effort to produce data flawed enough that our approach allows to discard it. But we believe this philosophy could become practically useful by adding more physical constraints, \textit{e.g.} with higher point correlators.

More generally, it should also be possible to generalize the framework to higher-point correlation functions and possibly to combine input from several such correlation functions to further constrain the spectral functions, and even access scattering data, using the positivity constraints familiar within the S-matrix bootstrap approach~\cite{Karateev:2019ymz}.

A natural next step is to study other models. Indeed, the numerical example we considered allowed to test the method on actual Monte-Carlo data, but this was for a rather simple quantum field theory with a sharp Dirac peak in the spectral function.  It would be interesting to test the method on phenomenologically more relevant problems, for instance the quark electromagnetic current-current correlator used both to determine the hadronic vacuum polarization and to reconstruct the $R$-ratio (this was previously studied using the modified Backus-Gilbert method \cite{ExtendedTwistedMassCollaborationETMC:2022sta}). Spectral reconstruction problems are also fundamental in condensed matter physics, and it would be interesting to see how this rigorous approach compares with the state of the art \cite{Jarrell:1996,bergeron2016,Levy_2017,PhysRevB.96.155128,PhysRevE.95.061302,fei2021,PhysRevLett.124.056401,KAUFMANN2023108519}.

Instead of reconstructing $\rho$, the semi-definite programming approach can also be used to access real-time data directly as was done in \cite{Lawrence:2024hjm,Mutzel:2025veg}. Here, this requires only bounding a non-convex real-time propagator with affine envelopes, which can be done simply by cutting it into convex/concave parts. 

Throughout this work, we kept the distributions $G$ acting on $\rho$ fixed (and in particular with fixed width). Intuitively, more precise Monte Carlo data allows to accurately bound distributions with narrower support. It would be interesting to see if, as with the Backus Gilbert method, there is a way to decide on an optimal width based on the data.

Finally, we have considered the task of estimating the spectral density associated to lattice data directly. If one aims to compute predictions for a continuum theory, one would then need to take a numerical continuum limit of our bounds, as $a\rightarrow 0$, with appropriate extrapolations. An alternative approach is to instead extrapolate the correlation functions first, and then use semi-definite programming to bound continuum spectral functions. It was shown how to deal with continuum theories with \emph{strict} (and not statistical) error bars  in \cite{Mutzel:2025veg}, but studying the extension to statistical (and systematic) error bars should be illuminating. An advantage of the continuum limit is that the spectral density is more constrained than in the discrete, and under mild conditions, it is exactly zero in finite intervals, between the excitation peaks corresponding to particles, and the multiparticle continuum. This allows to use a bootstrap approach, scanning over possible masses, the SDP being feasible only in a narrow range. We expect this method to be particularly powerful for the estimation of masses from Monte Carlo data.

\section*{Note}
\textit{As we were completing this article, which continues a line we developed in~\cite{Mutzel:2025veg}, a related and independent work by Abbott, Fields, Jay, Oare, and Saccardi appeared~\cite{abbott2026causalbootstrapboundingsmeared}, addressing a similar problem. Both works employ a Schur complement reduction to recast the original convex problem as an infinite-dimensional SDP, but the subsequent relaxations differ.}

\begin{acknowledgments}
We thank Adrìa Tort Martinez for discussions on the dual formulation. This project has received funding from the European Research Council (ERC) from the QFT.zip project (grant agreement No. 101040260).
\end{acknowledgments}

\bibliography{references}
\appendix

\section*{Appendix}
\section{Derivation of dual problem from Lagrange duality}\label{app:dualfromlagrangian}
In this appendix we derive the dual optimization problem \eqref{eq:dualSDP} from standard Lagrange duality. In the following we use the notation $(f_1,f_2)\equiv \int f_1(\omega)\,f_2(\omega) \; \d \omega$ for readability. Let us denote the optimal value of the primal problem \eqref{eq:primal} above by $p^*$. We now take the constraints into account by including them in the objective function\footnote{Note that all $\min$, $\max$ swap, if instead we want to maximize the objective function in the primal problem.}. 
We define the Lagrangian
\be
\cL(\rho; \nu, \Gamma)= (G, \rho) - (\nu, \rho)-\tr \l[\Gamma A(y(\rho))\r] \;,
\ee
where $\nu(\omega) \geq 0$ is a positive function (the Lagrange multiplier associated with the inequality constraint $\rho(\omega) \geq 0$) and $\Gamma$ is PSD (associated with the Källén-Lehmann constraint, \textit{i.e.}~the consistency with the Monte-Carlo data). 
Since A is affine in $\rho$ we can expand it
\be 
\begin{aligned}A(y(\rho))=
\underbrace{\left(\begin{array}{ll}
F_{\max} & -\bar{\cC}^T \\
-\bar{\cC} & \Sigma
\end{array}\right)}_{\cA_c} + (\underbrace{\left(\begin{array}{ll}
0 & \mathcal{K}(\omega)^T \\
\mathcal{K}(\omega) &0
\end{array}\right)}_{\cA(\omega)},\rho)
\end{aligned}
\ee
and hence
\be
\tr \l[\Gamma A(y)\r] = \tr (\Gamma \cA_c) + (\rho,\tr \l[\Gamma \cA(\omega)\r]) \;.
\ee
The Lagrange dual function is defined as the minimum value of the Lagrangian over $\rho(\omega)$:
\be
h(\nu,\Gamma) = \min_{\rho} \cL(\rho(\omega),\nu(\omega),\Gamma) \;.
\ee
The positivity of the Lagrange multipliers ensures that the dual function provides a lower bound on the primal optimum,
\be
h(\nu,\Gamma)\le p^*,
\ee
for any $\nu(\omega)\geq 0$ and $\Gamma\succeq 0$ \cite{boyd2004convex,NocedalWright}. The best such lower bound is obtained by maximizing the dual function over all admissible multipliers $\nu(\omega)$ and $\Gamma$, leading to the dual optimum
\be
d^*=\max_{\nu(\omega)\geq 0 ,\;\Gamma\succeq 0} f(\nu,\;\Gamma).
\ee 
Using the general min-max inequality
\be
\max_x \min_z f(x,z)
\le
\min_z \max_x f(x,z) 
\ee
and the definition of the dual function, we obtain
\be
\begin{split}
d^*&=\max_{\nu(\omega)\geq 0,\;\Gamma\succeq0} \min_\rho \cL(\rho,\nu,\Gamma)\\
&\leq
\min_\rho \max_{\nu(\omega)\geq 0,\;\Gamma\succeq 0} \cL(\rho,\nu,\Gamma) = p^*\;.
\end{split}
\ee
To determine the dual function explicitly, we minimize the Lagrangian over $\rho$. 
The minimum of the Lagrangian is easily found analytically, since the function is linear in $\rho$. 
The dual function is finite only if the coefficient of $\rho$ vanishes pointwise,
\be
\frac{\delta \cL}{\delta \rho}=G(\omega)-\nu(\omega)-\tr\!\left[\Gamma \cA(\omega)\right] \stackrel{!}{=} 0,
\quad
\forall \omega\in\mathbb R^+.
\ee
Otherwise the Lagrangian can be made arbitrarily negative and the dual function equals $-\infty$. 
Whenever this condition is satisfied, the dependence on $\rho$ drops out and the dual function reduces to
\be
h(\nu,\Gamma) = \min \cL = -\tr\!\left[\Gamma\cA_c\right].
\ee
Together with the positivity constraint $\nu(\omega)\geq 0$, we therefore obtain
\be
G(\omega) -\tr\!\left[\Gamma\cA(\omega)\right] \geq 0,
\qquad
\forall \omega\in\mathbb R^+,
\ee
and recover the dual optimization problem \eqref{eq:dualSDP}. 

\section{Dual formulation relaxation using bounds on propagator}
We now derive the dual formulation~\eqref{eq:dualrelaxed} of the relaxed primal problem~\eqref{eq:primal_slacked}, following the same Lagrangian construction as in the previous section but with the additional bracketing inequalities on the residuals $y_j$. 
The Lagrangian reads
\be\label{eq:Lrelaxed}
\begin{split}
\cL(\rho, y; \nu, \alpha, \beta, \Gamma)=&(G, \rho)-(\nu, \rho)-\tr[\Gamma A(y)]\\
&-\sum_j \alpha_j\left(\bar{C}_j+y_j-(f^-_{j}, \rho )\right)\\
&-\sum_j \beta_j\left(-y_j+ (f^+_{j}, \rho ) -\bar{C}_j\right)
\end{split}
\ee
with $\nu(\omega) \geq 0$ and $\Gamma \succeq 0$ as before, and additional non-negative multipliers $\alpha_j, \beta_j \geq 0$ associated respectively with the lower and upper bracketing inequalities \eqref{eq:inequality_relaxed_minus} and \eqref{eq:inequality_relaxed_plus}. 
The dual function is obtained by minimizing $\cL$ over the primal variables $\rho$ and $y$. Stationarity with respect to $y_j$ requires
\be\label{eq:stationarityxi}
\frac{\partial \cL}{\partial y_j} = -2\gamma_{0,j} - \alpha_j + \beta_j \stackrel{!}{=} 0 \;,
\ee
where we have used the parametrization~\eqref{eq:Gamma},
\be
\Gamma = \begin{pmatrix} \gamma_{00} & \gamma_0^T \\ 
\gamma_0 & \Gamma_\Sigma \end{pmatrix},
\ee
to identify $\partial \tr[\Gamma A(y)] / \partial y_j = 2\gamma_{0,j}$. Stationarity with respect to $\rho$ gives the pointwise condition
\be\label{eq:stationarityrho}
\frac{\delta \cL}{\delta \rho(\omega)} = G(\omega) - \nu(\omega) + \sum_j \alpha_j f_j^-(\omega) - \sum_j\beta_j f_j^+(\omega) \stackrel{!}{=} 0 \;.
\ee
Substituting~\eqref{eq:stationarityxi} and \eqref{eq:stationarityrho} back into the Lagrangian eliminates the $\rho$- and $y$-dependence. What remains is
\begin{align}
h(\nu, \Gamma, \alpha_j, \beta_j) =& \min_{\rho,y} \cL \\
=& -\gamma_{00} F_{\max} - \tr\l[\Gamma_\Sigma\, \Sigma \r] + \sum_j (\beta_j - \alpha_j)\, \bar{C}_j \nonumber
\end{align}
and using $\beta_j = \alpha_j+2\gamma_{0,j}$, the dual function takes the compact form $h = -\tr[\Gamma\,\cA_c]$.

The constraints follow from $\nu(\omega) \geq 0$ in~\eqref{eq:stationarityrho} 
and from the non-negativity and elimination of the multipliers. 
Eliminating $\beta_j$ in favour of $\alpha_j$ via~\eqref{eq:stationarityxi} converts $\beta_j \geq 0$ into 
$\alpha_j + 2\gamma_{0,j} \geq 0$, and \eqref{eq:stationarityrho} 
together with $\nu(\omega) \geq 0$ becomes the pointwise inequality
\begin{align}
G(\omega) + \sum_j \alpha_j f_j^-(\omega) 
- \sum_j (\alpha_j + &2\gamma_{0,j})\, f_j^+(\omega) \geq 0\;, \nonumber \\
&\forall \omega \in \mathbb{R}^+.
\end{align}
Collecting everything, the relaxed dual reads
\be
\begin{aligned}
\max_{\Gamma,\,\alpha}\quad & -\tr\bigl[\Gamma\,\cA_c\bigr] \\
\text{with}\quad 
& \cdot G(\omega) + \sum_j \alpha_j f_j^-(\omega) \\ 
& - \sum_j (\alpha_j + 2\gamma_{0,j})\, f_j^+(\omega) \geq 0\,,\qquad \forall \omega \in \mathbb{R}^+, \\
& \cdot \alpha_j \geq 0, \\
&\cdot \alpha_j + 2\gamma_{0,j} \geq 0, \\
& \cdot\Gamma \succeq 0,
\end{aligned}
\ee
recovering \eqref{eq:dualrelaxed}.
 
\section{Construction of the envelope coefficients}
\label{app:finitedimrelaxed}
In this appendix we explain how the envelope coefficients $f^\pm_{j,k}$ (piecewise constant case) and $f^{1,\pm}_{j,k}, f^{2,\pm}_{j,k}$ (piecewise affine case), appearing in the basis expansion~\eqref{eq:basisexpansion}, are precomputed from the kernel $K(\omega,t_j)$, and justify the reduction of the point-wise positivity constraint~\eqref{eq:positivityrelaxed} to a finite system of linear inequalities.

\paragraph{Piecewise constant envelopes.}
On each interval $[\omega_{k-1}, \omega_k]$ the upper and lower piecewise constant envelopes are the maximum and minimum of $K(\omega,t_j)$ on the interval. Since $K(\omega,t_j)$ is monotone decreasing on $\mathbb{R}^+$ (it is a sum of decaying exponentials), the extrema are attained at the endpoints,
\begin{equation}
f^+_{j,k} = K(\omega_{k-1},t_j), \qquad 
f^-_{j,k} = K(\omega_k,t_j).
\end{equation}

\paragraph{Piecewise affine envelopes.}
On each interval $[\omega_{k-1}, \omega_k]$ the upper envelope is the secant of $K(\omega,t_j)$,
\begin{equation}
f^+_j(\omega) = K(\omega_{k-1},t_j) + 
\frac{K(\omega_k,t_j) - K(\omega_{k-1},t_j)}{\omega_k - \omega_{k-1}}\,
(\omega - \omega_{k-1}),
\end{equation}
which lies above $K(\omega,t_j)$ on the interval by convexity. 
The lower envelope is the tangent at an interior point. We choose the midpoint $\bar\omega_k = (\omega_{k-1} + \omega_k)/2$ (an equally valid choice would be the geometric mean $\bar\omega_k = \sqrt{\omega_{k-1}\omega_k}$),
\begin{equation}
f^-_j(\omega) = K(\bar\omega_k,t_j) + 
K^\prime(\bar\omega_k,t_j)\,(\omega - \bar\omega_k),
\end{equation}
which lies below $K(\omega,t_j)$ on the interval, also by convexity. Identifying with the basis expansion, the precomputed coefficients are
\begin{equation}
\begin{split}
f^{1,+}_{j,k} &= K(\omega_{k-1},t_j)\,,\\
f^{2,+}_{j,k} &= \frac{K(\omega_k,t_j) - K(\omega_{k-1},t_j)}{\omega_k - \omega_{k-1}}\,,\\
f^{1,-}_{j,k} &= K(\bar\omega_k,t_j) - \frac{\partial K(\omega,t_j)}{\partial\omega}\vert_{\omega=\bar\omega_k}\,
(\bar\omega_k - \omega_{k-1})\,, \\
f^{2,-}_{j,k} &= \frac{\partial K(\omega,t_j)}{\partial\omega}\vert_{\omega=\bar\omega_k}\,.
\end{split}
\end{equation}
On the final unbounded interval $[\omega_{N_v}, \infty)$ the kernel 
decays exponentially. We use the constant envelopes $f^+_{j,N_v} = K(\omega_{N_v},t_j)$ and $f^-_{j>0,N_v} = 0$ (note that at $t_0=0$ the kernel $K(\omega,0)=1+e^{-T\omega}$ tends to $1$, not $0$, as $\omega \to \infty$ and thus we use $f^-_{j=0,N_v} = 1$ in this case). 

\paragraph{Reduction to endpoint inequalities.}
The reduction of the pointwise constraint~\eqref{eq:positivityrelaxed} to finitely many linear inequalities relies on a simple observation: a linear function of $\omega$ on a closed interval is non-negative everywhere on the interval if and only if it is non-negative at the two endpoints. The left-hand side of~\eqref{eq:positivityrelaxed}, restricted to a single interval, is linear in $\omega$ in the piecewise affine case (and constant in the piecewise constant case). Checking the two endpoints therefore certifies the inequality on the entire interval exactly. This yields the finite system of inequalities \eqref{eq:inequalities_affine} given in the main text.

The relaxed dual SDP is then
\begin{subequations}\label{eq:finiteSDP}
\begin{align}
\max_{\Gamma,\,\alpha} \quad & -\mathrm{tr}\bigl[\Gamma\,\mathcal{A}_c\bigr],\\
\text{under} \quad & \Gamma \succeq 0, 
\quad \alpha_j \geq 0, \quad \alpha_j + 2\gamma_{0,j} \geq 0,\\
& \text{the linear constraints }\eqref{eq:inequalities_affine} \text{ or } \eqref{eq:inequalities_constant},
\end{align}
\end{subequations}
which is a standard semidefinite program in the variables $\Gamma$ (a positive semidefinite $(N_c+1) \times (N_c+1)$ matrix) and $\alpha \in (\mathbb{R}^{+})^{N_c}$, solvable with any interior-point SDP solver.

\section{Regularization of covariance matrix}\label{app:regularization}

There is a subtlety in what one can mean by statistical consistency with the Monte-Carlo data. 
The constraint $y[\rho]^T M y[\rho] \leq F_{\max}$ in \eqref{eq:Fmaxconstraint} requires a choice of $M$. 
The natural choice is the inverse of the covariance matrix $M = \Sigma^{-1}$, and $F_{\max}$ computed on resampled binned data or resamples of the entire set of configurations. For Gaussian errors, this defines a $99\%$ confidence region. 

The estimation of the covariance matrix from a finite set of samples is however non-trivial.  
Given $N_{\rm config}$ statistically independent samples and $N_c$ correlator values, the sample covariance $\hat\Sigma$ has eigenvalue fluctuations of relative size $\sqrt{N_c/N_{\rm config}}$, as predicted by random matrix theory~\cite{Marcenko_1967}. In particular, small eigenvalues tend to be underestimated while large eigenvalues are overestimated. This issue is well known in lattice Monte-Carlo analyses~\cite{Yoon:2011wdw,Bruno:2022mfy}. 
As the sample size increases, $N_{\rm config} \gg N_c$, this correction becomes small.

A standard procedure is to regularize the small, poorly estimated eigenvalues of the covariance matrix. The simplest choice is to shift the eigenvalue spectrum by a small positive amount,
\be
\Sigma \rightarrow \Sigma_{\rm reg}
=
\Sigma+\epsilon \lVert \Sigma \rVert \mathbf{1}\;.
\ee
In our method, there are two places at which the covariance matrix $\Sigma$ enters, with different consequences for rigor:
\begin{itemize}
\item The covariance matrix appearing in $A(y(\rho))$ may always be replaced by $\Sigma + \epsilon \lVert \Sigma \rVert \mathbf{1}$ with $\epsilon \geq 0$ without compromising rigor: It amounts to replacing the constraint $A(y) \succeq 0$ by
$$
A(y)+E \succeq 0, 
$$
with 
$$
E=\left(\begin{array}{cc}
0 & 0 \\
0 & \epsilon \mathbf{1}
\end{array}\right) \succeq 0 \quad \text{for}  \quad \epsilon\geq 0 \;.
$$
Since $E$ is itself PSD, any feasible $\rho$ for the original problem automatically satisfies $A(y)+E \succeq 0$. Hence 
the regularized problem is a genuine relaxation of the original SDP.
\item Second, $\Sigma$ also enters on the right-hand side of Eq.~\eqref{eq:Fmaxconstr} through the distribution of the statistic $F_k$ used to determine $F_{\max}$, Eq.~\eqref{eq:Fk}. Regularizing $\Sigma$ in this step is \emph{not} a relaxation. It changes the empirical distribution of $F_k$ and hence the value of $F_{\max}$ itself. The resulting confidence region therefore corresponds to a different statistical test and is generally not directly comparable to the one obtained from the unregularized covariance matrix (in fact, it typically leads to a smaller value of $F_{\max}$ and thus to tighter constraints).
\end{itemize}
Regularizing $\Sigma$ in both the SDP constraint and in the determination of $F_{\max}$ defines a self-consistent confidence region, but for a different statistic. Such a procedure is not incorrect, but it no longer corresponds to a relaxation of the original problem. 
In our setup we therefore use the unregularized covariance throughout. 
If numerical conditioning of $\Sigma^{-1}$ becomes an issue, regularizing $\Sigma$ \emph{only} in $A(y)$ (and leaving $F_{\max}$ unregularized) is the rigorous choice. In the examples considered here the covariance matrix is sufficiently well conditioned.
\end{document}

%% file: macros.tex
\usepackage{amsmath}
\usepackage{amssymb}
\usepackage{graphicx} 
\usepackage{color}
\usepackage{braket}
\usepackage{enumitem}

\definecolor{linkcol}{RGB}{18, 94, 173}     
\definecolor{citecol}{RGB}{166, 45, 124}    
\definecolor{urlcol}{RGB}{0, 128, 112}      

\usepackage[
    colorlinks=true,
    linkcolor=linkcol,
    citecolor=citecol,
    urlcolor=urlcol,
]{hyperref}

\def\be{\begin{equation}}
\def\ee{\end{equation}}
\def\ba{\begin{align}}
\def\ea{\end{align}}
\def\bs{\begin{split}}
\def\es{\end{split}}
\def\bea{\begin{eqnarray}}
\def\eea{\end{eqnarray}}
\def\l{\left}
\def\r{\right}

\def\cA{\mathcal{A}}

\def\cC{\mathcal{C}}

\def\cL{\mathcal{L}}

\def\cC{\mathcal{C}}

\def\tr{\,\mathrm{tr}}

\def\xb{\mathbf{x}}
\def\yb{\mathbf{y}}

\definecolor{darkred}{rgb}{0.8,0.1,0.1}

\renewcommand{\d}{\mathrm{d}}
\newcommand{\upd}{\mathrm{d}}
\newcommand{\e}{\mathrm{e}}

%% file: main.bbl
\begin{thebibliography}{52}%
\makeatletter
\providecommand \@ifxundefined [1]{%
 \@ifx{#1\undefined}
}%
\providecommand \@ifnum [1]{%
 \ifnum #1\expandafter \@firstoftwo
 \else \expandafter \@secondoftwo
 \fi
}%
\providecommand \@ifx [1]{%
 \ifx #1\expandafter \@firstoftwo
 \else \expandafter \@secondoftwo
 \fi
}%
\providecommand \natexlab [1]{#1}%
\providecommand \enquote  [1]{``#1''}%
\providecommand \bibnamefont  [1]{#1}%
\providecommand \bibfnamefont [1]{#1}%
\providecommand \citenamefont [1]{#1}%
\providecommand \href@noop [0]{\@secondoftwo}%
\providecommand \href [0]{\begingroup \@sanitize@url \@href}%
\providecommand \@href[1]{\@@startlink{#1}\@@href}%
\providecommand \@@href[1]{\endgroup#1\@@endlink}%
\providecommand \@sanitize@url [0]{\catcode `\\12\catcode `\$12\catcode
  `\&12\catcode `\#12\catcode `\^12\catcode `\_12\catcode `\%12\relax}%
\providecommand \@@startlink[1]{}%
\providecommand \@@endlink[0]{}%
\providecommand \url  [0]{\begingroup\@sanitize@url \@url }%
\providecommand \@url [1]{\endgroup\@href {#1}{\urlprefix }}%
\providecommand \urlprefix  [0]{URL }%
\providecommand \Eprint [0]{\href }%
\providecommand \doibase [0]{https://doi.org/}%
\providecommand \selectlanguage [0]{\@gobble}%
\providecommand \bibinfo  [0]{\@secondoftwo}%
\providecommand \bibfield  [0]{\@secondoftwo}%
\providecommand \translation [1]{[#1]}%
\providecommand \BibitemOpen [0]{}%
\providecommand \bibitemStop [0]{}%
\providecommand \bibitemNoStop [0]{.\EOS\space}%
\providecommand \EOS [0]{\spacefactor3000\relax}%
\providecommand \BibitemShut  [1]{\csname bibitem#1\endcsname}%
\let\auto@bib@innerbib\@empty
\bibitem [{\citenamefont {Durr}\ \emph {et~al.}(2008)\citenamefont {Durr} \emph
  {et~al.}}]{BMW:2008jgk}%
  \BibitemOpen
  \bibfield  {author} {\bibinfo {author} {\bibfnamefont {S.}~\bibnamefont
  {Durr}} \emph {et~al.} (\bibinfo {collaboration} {BMW}),\ }\bibfield  {title}
  {\bibinfo {title} {{Ab-Initio Determination of Light Hadron Masses}},\ }\href
  {https://doi.org/10.1126/science.1163233} {\bibfield  {journal} {\bibinfo
  {journal} {Science}\ }\textbf {\bibinfo {volume} {322}},\ \bibinfo {pages}
  {1224} (\bibinfo {year} {2008})},\ \Eprint {https://arxiv.org/abs/0906.3599}
  {arXiv:0906.3599 [hep-lat]} \BibitemShut {NoStop}%
\bibitem [{\citenamefont {Aoki}\ \emph {et~al.}(2026)\citenamefont {Aoki} \emph
  {et~al.}}]{FlavourLatticeAveragingGroupFLAG:2024oxs}%
  \BibitemOpen
  \bibfield  {author} {\bibinfo {author} {\bibfnamefont {Y.}~\bibnamefont
  {Aoki}} \emph {et~al.} (\bibinfo {collaboration} {Flavour Lattice Averaging
  Group (FLAG)}),\ }\bibfield  {title} {\bibinfo {title} {{FLAG review 2024}},\
  }\href {https://doi.org/10.1103/nfzp-p5dn} {\bibfield  {journal} {\bibinfo
  {journal} {Phys. Rev. D}\ }\textbf {\bibinfo {volume} {113}},\ \bibinfo
  {pages} {014508} (\bibinfo {year} {2026})},\ \Eprint
  {https://arxiv.org/abs/2411.04268} {arXiv:2411.04268 [hep-lat]} \BibitemShut
  {NoStop}%
\bibitem [{\citenamefont {Boccaletti}\ \emph {et~al.}(2026)\citenamefont
  {Boccaletti} \emph {et~al.}}]{Boccaletti:2024guq}%
  \BibitemOpen
  \bibfield  {author} {\bibinfo {author} {\bibfnamefont {A.}~\bibnamefont
  {Boccaletti}} \emph {et~al.},\ }\bibfield  {title} {\bibinfo {title} {{Hybrid
  calculation of hadronic vacuum polarization in muon g {\ensuremath{-}} 2 to
  0.48{\%}}},\ }\href {https://doi.org/10.1038/s41586-026-10449-z} {\bibfield
  {journal} {\bibinfo  {journal} {Nature}\ }\textbf {\bibinfo {volume} {653}},\
  \bibinfo {pages} {373} (\bibinfo {year} {2026})},\ \Eprint
  {https://arxiv.org/abs/2407.10913} {arXiv:2407.10913 [hep-lat]} \BibitemShut
  {NoStop}%
\bibitem [{\citenamefont {Jarrell}\ and\ \citenamefont
  {Gubernatis}(1996)}]{Jarrell:1996}%
  \BibitemOpen
  \bibfield  {author} {\bibinfo {author} {\bibfnamefont {M.}~\bibnamefont
  {Jarrell}}\ and\ \bibinfo {author} {\bibfnamefont {J.}~\bibnamefont
  {Gubernatis}},\ }\bibfield  {title} {\bibinfo {title} {Bayesian inference and
  the analytic continuation of imaginary-time quantum monte carlo data},\
  }\href {https://doi.org/https://doi.org/10.1016/0370-1573(95)00074-7}
  {\bibfield  {journal} {\bibinfo  {journal} {Physics Reports}\ }\textbf
  {\bibinfo {volume} {269}},\ \bibinfo {pages} {133} (\bibinfo {year}
  {1996})}\BibitemShut {NoStop}%
\bibitem [{\citenamefont {Asakawa}\ \emph {et~al.}(2001)\citenamefont
  {Asakawa}, \citenamefont {Hatsuda},\ and\ \citenamefont
  {Nakahara}}]{Asakawa:2000tr}%
  \BibitemOpen
  \bibfield  {author} {\bibinfo {author} {\bibfnamefont {M.}~\bibnamefont
  {Asakawa}}, \bibinfo {author} {\bibfnamefont {T.}~\bibnamefont {Hatsuda}},\
  and\ \bibinfo {author} {\bibfnamefont {Y.}~\bibnamefont {Nakahara}},\
  }\bibfield  {title} {\bibinfo {title} {{Maximum entropy analysis of the
  spectral functions in lattice QCD}},\ }\href
  {https://doi.org/10.1016/S0146-6410(01)00150-8} {\bibfield  {journal}
  {\bibinfo  {journal} {Prog. Part. Nucl. Phys.}\ }\textbf {\bibinfo {volume}
  {46}},\ \bibinfo {pages} {459} (\bibinfo {year} {2001})},\ \Eprint
  {https://arxiv.org/abs/hep-lat/0011040} {arXiv:hep-lat/0011040} \BibitemShut
  {NoStop}%
\bibitem [{\citenamefont {Burnier}\ and\ \citenamefont
  {Rothkopf}(2013)}]{PhysRevLett.111.182003}%
  \BibitemOpen
  \bibfield  {author} {\bibinfo {author} {\bibfnamefont {Y.}~\bibnamefont
  {Burnier}}\ and\ \bibinfo {author} {\bibfnamefont {A.}~\bibnamefont
  {Rothkopf}},\ }\bibfield  {title} {\bibinfo {title} {Bayesian approach to
  spectral function reconstruction for euclidean quantum field theories},\
  }\href {https://doi.org/10.1103/PhysRevLett.111.182003} {\bibfield  {journal}
  {\bibinfo  {journal} {Phys. Rev. Lett.}\ }\textbf {\bibinfo {volume} {111}},\
  \bibinfo {pages} {182003} (\bibinfo {year} {2013})}\BibitemShut {NoStop}%
\bibitem [{\citenamefont
  {Beach}(2004)}]{beach2004identifyingmaximumentropymethod}%
  \BibitemOpen
  \bibfield  {author} {\bibinfo {author} {\bibfnamefont {K.~S.~D.}\
  \bibnamefont {Beach}},\ }\href {https://arxiv.org/abs/cond-mat/0403055}
  {\bibinfo {title} {Identifying the maximum entropy method as a special limit
  of stochastic analytic continuation}} (\bibinfo {year} {2004}),\ \Eprint
  {https://arxiv.org/abs/cond-mat/0403055} {arXiv:cond-mat/0403055
  [cond-mat.str-el]} \BibitemShut {NoStop}%
\bibitem [{\citenamefont {Rothkopf}(2022)}]{Rothkopf:2022ctl}%
  \BibitemOpen
  \bibfield  {author} {\bibinfo {author} {\bibfnamefont {A.}~\bibnamefont
  {Rothkopf}},\ }\bibfield  {title} {\bibinfo {title} {{Bayesian inference of
  real-time dynamics from lattice QCD}},\ }\href
  {https://doi.org/10.3389/fphy.2022.1028995} {\bibfield  {journal} {\bibinfo
  {journal} {Front. Phys.}\ }\textbf {\bibinfo {volume} {10}},\ \bibinfo
  {pages} {1028995} (\bibinfo {year} {2022})},\ \Eprint
  {https://arxiv.org/abs/2208.13590} {arXiv:2208.13590 [hep-lat]} \BibitemShut
  {NoStop}%
\bibitem [{\citenamefont {Rothkopf}(2017)}]{PhysRevD.95.056016}%
  \BibitemOpen
  \bibfield  {author} {\bibinfo {author} {\bibfnamefont {A.}~\bibnamefont
  {Rothkopf}},\ }\bibfield  {title} {\bibinfo {title} {Bayesian inference of
  nonpositive spectral functions in quantum field theory},\ }\href
  {https://doi.org/10.1103/PhysRevD.95.056016} {\bibfield  {journal} {\bibinfo
  {journal} {Phys. Rev. D}\ }\textbf {\bibinfo {volume} {95}},\ \bibinfo
  {pages} {056016} (\bibinfo {year} {2017})}\BibitemShut {NoStop}%
\bibitem [{\citenamefont {Backus}\ and\ \citenamefont
  {Gilbert}(1970)}]{BackusGilbert}%
  \BibitemOpen
  \bibfield  {author} {\bibinfo {author} {\bibfnamefont {G.}~\bibnamefont
  {Backus}}\ and\ \bibinfo {author} {\bibfnamefont {F.}~\bibnamefont
  {Gilbert}},\ }\bibfield  {title} {\bibinfo {title} {Uniqueness in the
  inversion of inaccurate gross earth data},\ }\href
  {http://www.jstor.org/stable/73746} {\bibfield  {journal} {\bibinfo
  {journal} {Philosophical Transactions of the Royal Society of London. Series
  A, Mathematical and Physical Sciences}\ }\textbf {\bibinfo {volume} {266}},\
  \bibinfo {pages} {123} (\bibinfo {year} {1970})}\BibitemShut {NoStop}%
\bibitem [{\citenamefont {Hansen}\ \emph {et~al.}(2017)\citenamefont {Hansen},
  \citenamefont {Meyer},\ and\ \citenamefont {Robaina}}]{Hansen:2017mnd}%
  \BibitemOpen
  \bibfield  {author} {\bibinfo {author} {\bibfnamefont {M.~T.}\ \bibnamefont
  {Hansen}}, \bibinfo {author} {\bibfnamefont {H.~B.}\ \bibnamefont {Meyer}},\
  and\ \bibinfo {author} {\bibfnamefont {D.}~\bibnamefont {Robaina}},\
  }\bibfield  {title} {\bibinfo {title} {{From deep inelastic scattering to
  heavy-flavor semileptonic decays: Total rates into multihadron final states
  from lattice QCD}},\ }\href {https://doi.org/10.1103/PhysRevD.96.094513}
  {\bibfield  {journal} {\bibinfo  {journal} {Phys. Rev. D}\ }\textbf {\bibinfo
  {volume} {96}},\ \bibinfo {pages} {094513} (\bibinfo {year} {2017})},\
  \Eprint {https://arxiv.org/abs/1704.08993} {arXiv:1704.08993 [hep-lat]}
  \BibitemShut {NoStop}%
\bibitem [{\citenamefont {Hansen}\ \emph {et~al.}(2019)\citenamefont {Hansen},
  \citenamefont {Lupo},\ and\ \citenamefont {Tantalo}}]{Hansen:2019idp}%
  \BibitemOpen
  \bibfield  {author} {\bibinfo {author} {\bibfnamefont {M.}~\bibnamefont
  {Hansen}}, \bibinfo {author} {\bibfnamefont {A.}~\bibnamefont {Lupo}},\ and\
  \bibinfo {author} {\bibfnamefont {N.}~\bibnamefont {Tantalo}},\ }\bibfield
  {title} {\bibinfo {title} {{Extraction of spectral densities from lattice
  correlators}},\ }\href {https://doi.org/10.1103/PhysRevD.99.094508}
  {\bibfield  {journal} {\bibinfo  {journal} {Phys. Rev. D}\ }\textbf {\bibinfo
  {volume} {99}},\ \bibinfo {pages} {094508} (\bibinfo {year} {2019})},\
  \Eprint {https://arxiv.org/abs/1903.06476} {arXiv:1903.06476 [hep-lat]}
  \BibitemShut {NoStop}%
\bibitem [{\citenamefont {Tripolt}\ \emph {et~al.}(2019)\citenamefont
  {Tripolt}, \citenamefont {Gubler}, \citenamefont {Ulybyshev},\ and\
  \citenamefont {Von~Smekal}}]{Tripolt:2018xeo}%
  \BibitemOpen
  \bibfield  {author} {\bibinfo {author} {\bibfnamefont {R.-A.}\ \bibnamefont
  {Tripolt}}, \bibinfo {author} {\bibfnamefont {P.}~\bibnamefont {Gubler}},
  \bibinfo {author} {\bibfnamefont {M.}~\bibnamefont {Ulybyshev}},\ and\
  \bibinfo {author} {\bibfnamefont {L.}~\bibnamefont {Von~Smekal}},\ }\bibfield
   {title} {\bibinfo {title} {{Numerical analytic continuation of Euclidean
  data}},\ }\href {https://doi.org/10.1016/j.cpc.2018.11.012} {\bibfield
  {journal} {\bibinfo  {journal} {Comput. Phys. Commun.}\ }\textbf {\bibinfo
  {volume} {237}},\ \bibinfo {pages} {129} (\bibinfo {year} {2019})},\ \Eprint
  {https://arxiv.org/abs/1801.10348} {arXiv:1801.10348 [hep-ph]} \BibitemShut
  {NoStop}%
\bibitem [{\citenamefont {Schlessinger}(1968)}]{PhysRev.167.1411}%
  \BibitemOpen
  \bibfield  {author} {\bibinfo {author} {\bibfnamefont {L.}~\bibnamefont
  {Schlessinger}},\ }\bibfield  {title} {\bibinfo {title} {Use of analyticity
  in the calculation of nonrelativistic scattering amplitudes},\ }\href
  {https://doi.org/10.1103/PhysRev.167.1411} {\bibfield  {journal} {\bibinfo
  {journal} {Phys. Rev.}\ }\textbf {\bibinfo {volume} {167}},\ \bibinfo {pages}
  {1411} (\bibinfo {year} {1968})}\BibitemShut {NoStop}%
\bibitem [{\citenamefont {Bergamaschi}\ \emph {et~al.}(2023)\citenamefont
  {Bergamaschi}, \citenamefont {Jay},\ and\ \citenamefont
  {Oare}}]{Bergamaschi:2023xzx}%
  \BibitemOpen
  \bibfield  {author} {\bibinfo {author} {\bibfnamefont {T.}~\bibnamefont
  {Bergamaschi}}, \bibinfo {author} {\bibfnamefont {W.~I.}\ \bibnamefont
  {Jay}},\ and\ \bibinfo {author} {\bibfnamefont {P.~R.}\ \bibnamefont
  {Oare}},\ }\bibfield  {title} {\bibinfo {title} {{Hadronic structure,
  conformal maps, and analytic continuation}},\ }\href
  {https://doi.org/10.1103/PhysRevD.108.074516} {\bibfield  {journal} {\bibinfo
   {journal} {Phys. Rev. D}\ }\textbf {\bibinfo {volume} {108}},\ \bibinfo
  {pages} {074516} (\bibinfo {year} {2023})},\ \Eprint
  {https://arxiv.org/abs/2305.16190} {arXiv:2305.16190 [hep-lat]} \BibitemShut
  {NoStop}%
\bibitem [{\citenamefont {Sandvik}(1998)}]{PhysRevB.57.10287}%
  \BibitemOpen
  \bibfield  {author} {\bibinfo {author} {\bibfnamefont {A.~W.}\ \bibnamefont
  {Sandvik}},\ }\bibfield  {title} {\bibinfo {title} {Stochastic method for
  analytic continuation of quantum monte carlo data},\ }\href
  {https://doi.org/10.1103/PhysRevB.57.10287} {\bibfield  {journal} {\bibinfo
  {journal} {Phys. Rev. B}\ }\textbf {\bibinfo {volume} {57}},\ \bibinfo
  {pages} {10287} (\bibinfo {year} {1998})}\BibitemShut {NoStop}%
\bibitem [{\citenamefont {Buzzicotti}\ \emph {et~al.}(2024)\citenamefont
  {Buzzicotti}, \citenamefont {De~Santis},\ and\ \citenamefont
  {Tantalo}}]{Buzzicotti:2023qdv}%
  \BibitemOpen
  \bibfield  {author} {\bibinfo {author} {\bibfnamefont {M.}~\bibnamefont
  {Buzzicotti}}, \bibinfo {author} {\bibfnamefont {A.}~\bibnamefont
  {De~Santis}},\ and\ \bibinfo {author} {\bibfnamefont {N.}~\bibnamefont
  {Tantalo}},\ }\bibfield  {title} {\bibinfo {title} {{Teaching to extract
  spectral densities from lattice correlators to a broad audience of
  learning-machines}},\ }\href
  {https://doi.org/10.1140/epjc/s10052-024-12399-0} {\bibfield  {journal}
  {\bibinfo  {journal} {Eur. Phys. J. C}\ }\textbf {\bibinfo {volume} {84}},\
  \bibinfo {pages} {32} (\bibinfo {year} {2024})},\ \Eprint
  {https://arxiv.org/abs/2307.00808} {arXiv:2307.00808 [hep-lat]} \BibitemShut
  {NoStop}%
\bibitem [{\citenamefont {Wang}\ \emph {et~al.}(2022)\citenamefont {Wang},
  \citenamefont {Shi},\ and\ \citenamefont {Zhou}}]{Wang:2021jou}%
  \BibitemOpen
  \bibfield  {author} {\bibinfo {author} {\bibfnamefont {L.}~\bibnamefont
  {Wang}}, \bibinfo {author} {\bibfnamefont {S.}~\bibnamefont {Shi}},\ and\
  \bibinfo {author} {\bibfnamefont {K.}~\bibnamefont {Zhou}},\ }\bibfield
  {title} {\bibinfo {title} {{Reconstructing spectral functions via automatic
  differentiation}},\ }\href {https://doi.org/10.1103/PhysRevD.106.L051502}
  {\bibfield  {journal} {\bibinfo  {journal} {Phys. Rev. D}\ }\textbf {\bibinfo
  {volume} {106}},\ \bibinfo {pages} {L051502} (\bibinfo {year} {2022})},\
  \Eprint {https://arxiv.org/abs/2111.14760} {arXiv:2111.14760 [hep-ph]}
  \BibitemShut {NoStop}%
\bibitem [{\citenamefont {Fournier}\ \emph {et~al.}(2020)\citenamefont
  {Fournier}, \citenamefont {Wang}, \citenamefont {Yazyev},\ and\ \citenamefont
  {Wu}}]{PhysRevLett.124.056401}%
  \BibitemOpen
  \bibfield  {author} {\bibinfo {author} {\bibfnamefont {R.}~\bibnamefont
  {Fournier}}, \bibinfo {author} {\bibfnamefont {L.}~\bibnamefont {Wang}},
  \bibinfo {author} {\bibfnamefont {O.~V.}\ \bibnamefont {Yazyev}},\ and\
  \bibinfo {author} {\bibfnamefont {Q.}~\bibnamefont {Wu}},\ }\bibfield
  {title} {\bibinfo {title} {Artificial neural network approach to the analytic
  continuation problem},\ }\href
  {https://doi.org/10.1103/PhysRevLett.124.056401} {\bibfield  {journal}
  {\bibinfo  {journal} {Phys. Rev. Lett.}\ }\textbf {\bibinfo {volume} {124}},\
  \bibinfo {pages} {056401} (\bibinfo {year} {2020})}\BibitemShut {NoStop}%
\bibitem [{\citenamefont {Horak}\ \emph {et~al.}(2022)\citenamefont {Horak},
  \citenamefont {Pawlowski}, \citenamefont {Rodr{\'\i}guez-Quintero},
  \citenamefont {Turnwald}, \citenamefont {Urban}, \citenamefont {Wink},\ and\
  \citenamefont {Zafeiropoulos}}]{Horak:2021syv}%
  \BibitemOpen
  \bibfield  {author} {\bibinfo {author} {\bibfnamefont {J.}~\bibnamefont
  {Horak}}, \bibinfo {author} {\bibfnamefont {J.~M.}\ \bibnamefont
  {Pawlowski}}, \bibinfo {author} {\bibfnamefont {J.}~\bibnamefont
  {Rodr{\'\i}guez-Quintero}}, \bibinfo {author} {\bibfnamefont
  {J.}~\bibnamefont {Turnwald}}, \bibinfo {author} {\bibfnamefont {J.~M.}\
  \bibnamefont {Urban}}, \bibinfo {author} {\bibfnamefont {N.}~\bibnamefont
  {Wink}},\ and\ \bibinfo {author} {\bibfnamefont {S.}~\bibnamefont
  {Zafeiropoulos}},\ }\bibfield  {title} {\bibinfo {title} {{Reconstructing QCD
  spectral functions with Gaussian processes}},\ }\href
  {https://doi.org/10.1103/PhysRevD.105.036014} {\bibfield  {journal} {\bibinfo
   {journal} {Phys. Rev. D}\ }\textbf {\bibinfo {volume} {105}},\ \bibinfo
  {pages} {036014} (\bibinfo {year} {2022})},\ \Eprint
  {https://arxiv.org/abs/2107.13464} {arXiv:2107.13464 [hep-ph]} \BibitemShut
  {NoStop}%
\bibitem [{\citenamefont {Bruno}\ \emph {et~al.}(2025)\citenamefont {Bruno},
  \citenamefont {Giusti},\ and\ \citenamefont {Saccardi}}]{Bruno:2024fqc}%
  \BibitemOpen
  \bibfield  {author} {\bibinfo {author} {\bibfnamefont {M.}~\bibnamefont
  {Bruno}}, \bibinfo {author} {\bibfnamefont {L.}~\bibnamefont {Giusti}},\ and\
  \bibinfo {author} {\bibfnamefont {M.}~\bibnamefont {Saccardi}},\ }\bibfield
  {title} {\bibinfo {title} {{Spectral densities from Euclidean lattice
  correlators via the Mellin transform}},\ }\href
  {https://doi.org/10.1103/PhysRevD.111.094515} {\bibfield  {journal} {\bibinfo
   {journal} {Phys. Rev. D}\ }\textbf {\bibinfo {volume} {111}},\ \bibinfo
  {pages} {094515} (\bibinfo {year} {2025})},\ \Eprint
  {https://arxiv.org/abs/2407.04141} {arXiv:2407.04141 [hep-lat]} \BibitemShut
  {NoStop}%
\bibitem [{\citenamefont {Lawrence}(2024)}]{Lawrence:2024hjm}%
  \BibitemOpen
  \bibfield  {author} {\bibinfo {author} {\bibfnamefont {S.}~\bibnamefont
  {Lawrence}},\ }\bibfield  {title} {\bibinfo {title} {{Model-free spectral
  reconstruction via Lagrange duality}},\ }\Eprint
  {https://arxiv.org/abs/2408.11766} {arXiv:2408.11766 [hep-lat]}  (\bibinfo
  {year} {2024})\BibitemShut {NoStop}%
\bibitem [{\citenamefont {Rattazzi}\ \emph {et~al.}(2008)\citenamefont
  {Rattazzi}, \citenamefont {Rychkov}, \citenamefont {Tonni},\ and\
  \citenamefont {Vichi}}]{Rattazzi:2008pe}%
  \BibitemOpen
  \bibfield  {author} {\bibinfo {author} {\bibfnamefont {R.}~\bibnamefont
  {Rattazzi}}, \bibinfo {author} {\bibfnamefont {V.~S.}\ \bibnamefont
  {Rychkov}}, \bibinfo {author} {\bibfnamefont {E.}~\bibnamefont {Tonni}},\
  and\ \bibinfo {author} {\bibfnamefont {A.}~\bibnamefont {Vichi}},\ }\bibfield
   {title} {\bibinfo {title} {{Bounding scalar operator dimensions in 4D
  CFT}},\ }\href {https://doi.org/10.1088/1126-6708/2008/12/031} {\bibfield
  {journal} {\bibinfo  {journal} {JHEP}\ }\textbf {\bibinfo {volume} {12}},\
  \bibinfo {pages} {031}},\ \Eprint {https://arxiv.org/abs/0807.0004}
  {arXiv:0807.0004 [hep-th]} \BibitemShut {NoStop}%
\bibitem [{\citenamefont {Poland}\ \emph {et~al.}(2019)\citenamefont {Poland},
  \citenamefont {Rychkov},\ and\ \citenamefont {Vichi}}]{Poland:2018epd}%
  \BibitemOpen
  \bibfield  {author} {\bibinfo {author} {\bibfnamefont {D.}~\bibnamefont
  {Poland}}, \bibinfo {author} {\bibfnamefont {S.}~\bibnamefont {Rychkov}},\
  and\ \bibinfo {author} {\bibfnamefont {A.}~\bibnamefont {Vichi}},\ }\bibfield
   {title} {\bibinfo {title} {{The Conformal Bootstrap: Theory, Numerical
  Techniques, and Applications}},\ }\href
  {https://doi.org/10.1103/RevModPhys.91.015002} {\bibfield  {journal}
  {\bibinfo  {journal} {Rev. Mod. Phys.}\ }\textbf {\bibinfo {volume} {91}},\
  \bibinfo {pages} {015002} (\bibinfo {year} {2019})},\ \Eprint
  {https://arxiv.org/abs/1805.04405} {arXiv:1805.04405 [hep-th]} \BibitemShut
  {NoStop}%
\bibitem [{\citenamefont {Paulos}\ \emph {et~al.}(2017)\citenamefont {Paulos},
  \citenamefont {Penedones}, \citenamefont {Toledo}, \citenamefont {van Rees},\
  and\ \citenamefont {Vieira}}]{Paulos:2016fap}%
  \BibitemOpen
  \bibfield  {author} {\bibinfo {author} {\bibfnamefont {M.~F.}\ \bibnamefont
  {Paulos}}, \bibinfo {author} {\bibfnamefont {J.}~\bibnamefont {Penedones}},
  \bibinfo {author} {\bibfnamefont {J.}~\bibnamefont {Toledo}}, \bibinfo
  {author} {\bibfnamefont {B.~C.}\ \bibnamefont {van Rees}},\ and\ \bibinfo
  {author} {\bibfnamefont {P.}~\bibnamefont {Vieira}},\ }\bibfield  {title}
  {\bibinfo {title} {{The S-matrix bootstrap. Part I: QFT in AdS}},\ }\href
  {https://doi.org/10.1007/JHEP11(2017)133} {\bibfield  {journal} {\bibinfo
  {journal} {JHEP}\ }\textbf {\bibinfo {volume} {11}},\ \bibinfo {pages}
  {133}},\ \Eprint {https://arxiv.org/abs/1607.06109} {arXiv:1607.06109
  [hep-th]} \BibitemShut {NoStop}%
\bibitem [{\citenamefont {Kruczenski}\ \emph {et~al.}(2022)\citenamefont
  {Kruczenski}, \citenamefont {Penedones},\ and\ \citenamefont {van
  Rees}}]{Kruczenski:2022lot}%
  \BibitemOpen
  \bibfield  {author} {\bibinfo {author} {\bibfnamefont {M.}~\bibnamefont
  {Kruczenski}}, \bibinfo {author} {\bibfnamefont {J.}~\bibnamefont
  {Penedones}},\ and\ \bibinfo {author} {\bibfnamefont {B.~C.}\ \bibnamefont
  {van Rees}},\ }\bibfield  {title} {\bibinfo {title} {{Snowmass White Paper:
  S-matrix Bootstrap}},\ }\Eprint {https://arxiv.org/abs/2203.02421}
  {arXiv:2203.02421 [hep-th]}  (\bibinfo {year} {2022})\BibitemShut {NoStop}%
\bibitem [{\citenamefont {Karateev}\ \emph {et~al.}(2020)\citenamefont
  {Karateev}, \citenamefont {Kuhn},\ and\ \citenamefont
  {Penedones}}]{Karateev:2019ymz}%
  \BibitemOpen
  \bibfield  {author} {\bibinfo {author} {\bibfnamefont {D.}~\bibnamefont
  {Karateev}}, \bibinfo {author} {\bibfnamefont {S.}~\bibnamefont {Kuhn}},\
  and\ \bibinfo {author} {\bibfnamefont {J.~a.}\ \bibnamefont {Penedones}},\
  }\bibfield  {title} {\bibinfo {title} {{Bootstrapping Massive Quantum Field
  Theories}},\ }\href {https://doi.org/10.1007/JHEP07(2020)035} {\bibfield
  {journal} {\bibinfo  {journal} {JHEP}\ }\textbf {\bibinfo {volume} {07}},\
  \bibinfo {pages} {035}},\ \Eprint {https://arxiv.org/abs/1912.08940}
  {arXiv:1912.08940 [hep-th]} \BibitemShut {NoStop}%
\bibitem [{\citenamefont {Chen}\ \emph {et~al.}(2022)\citenamefont {Chen},
  \citenamefont {Fitzpatrick},\ and\ \citenamefont {Karateev}}]{Chen:2021pgx}%
  \BibitemOpen
  \bibfield  {author} {\bibinfo {author} {\bibfnamefont {H.}~\bibnamefont
  {Chen}}, \bibinfo {author} {\bibfnamefont {A.~L.}\ \bibnamefont
  {Fitzpatrick}},\ and\ \bibinfo {author} {\bibfnamefont {D.}~\bibnamefont
  {Karateev}},\ }\bibfield  {title} {\bibinfo {title} {{Bootstrapping 2d
  \ensuremath{\phi}$^{4}$ theory with Hamiltonian truncation data}},\ }\href
  {https://doi.org/10.1007/JHEP02(2022)146} {\bibfield  {journal} {\bibinfo
  {journal} {JHEP}\ }\textbf {\bibinfo {volume} {02}},\ \bibinfo {pages}
  {146}},\ \Eprint {https://arxiv.org/abs/2107.10286} {arXiv:2107.10286
  [hep-th]} \BibitemShut {NoStop}%
\bibitem [{\citenamefont {Osterwalder}\ and\ \citenamefont
  {Schrader}(1975)}]{Osterwalder:1974tc}%
  \BibitemOpen
  \bibfield  {author} {\bibinfo {author} {\bibfnamefont {K.}~\bibnamefont
  {Osterwalder}}\ and\ \bibinfo {author} {\bibfnamefont {R.}~\bibnamefont
  {Schrader}},\ }\bibfield  {title} {\bibinfo {title} {{Axioms for Euclidean
  Green's Functions. 2.}},\ }\href {https://doi.org/10.1007/BF01608978}
  {\bibfield  {journal} {\bibinfo  {journal} {Commun. Math. Phys.}\ }\textbf
  {\bibinfo {volume} {42}},\ \bibinfo {pages} {281} (\bibinfo {year}
  {1975})}\BibitemShut {NoStop}%
\bibitem [{\citenamefont {Montvay}\ and\ \citenamefont
  {M{\"u}nster}(1994)}]{montvay1994quantum}%
  \BibitemOpen
  \bibfield  {author} {\bibinfo {author} {\bibfnamefont {I.}~\bibnamefont
  {Montvay}}\ and\ \bibinfo {author} {\bibfnamefont {G.}~\bibnamefont
  {M{\"u}nster}},\ }\href {https://books.google.fr/books?id=NHZshmEBXhcC}
  {\emph {\bibinfo {title} {Quantum Fields on a Lattice}}},\ Cambridge
  Monographs on Mathematical Physics\ (\bibinfo  {publisher} {Cambridge
  University Press},\ \bibinfo {year} {1994})\BibitemShut {NoStop}%
\bibitem [{\citenamefont {Delcamp}\ and\ \citenamefont
  {Tilloy}(2020)}]{Delcamp:2020hzo}%
  \BibitemOpen
  \bibfield  {author} {\bibinfo {author} {\bibfnamefont {C.}~\bibnamefont
  {Delcamp}}\ and\ \bibinfo {author} {\bibfnamefont {A.}~\bibnamefont
  {Tilloy}},\ }\bibfield  {title} {\bibinfo {title} {{Computing the
  renormalization group flow of two-dimensional $\phi^4$ theory with tensor
  networks}},\ }\href {https://doi.org/10.1103/PhysRevResearch.2.033278}
  {\bibfield  {journal} {\bibinfo  {journal} {Phys. Rev. Res.}\ }\textbf
  {\bibinfo {volume} {2}},\ \bibinfo {pages} {033278} (\bibinfo {year}
  {2020})},\ \Eprint {https://arxiv.org/abs/2003.12993} {arXiv:2003.12993
  [cond-mat.str-el]} \BibitemShut {NoStop}%
\bibitem [{\citenamefont {Loinaz}\ and\ \citenamefont
  {Willey}(1998)}]{Loinaz:1997az}%
  \BibitemOpen
  \bibfield  {author} {\bibinfo {author} {\bibfnamefont {W.}~\bibnamefont
  {Loinaz}}\ and\ \bibinfo {author} {\bibfnamefont {R.~S.}\ \bibnamefont
  {Willey}},\ }\bibfield  {title} {\bibinfo {title} {{Monte Carlo simulation
  calculation of critical coupling constant for continuum phi**4 in
  two-dimensions}},\ }\href {https://doi.org/10.1103/PhysRevD.58.076003}
  {\bibfield  {journal} {\bibinfo  {journal} {Phys. Rev. D}\ }\textbf {\bibinfo
  {volume} {58}},\ \bibinfo {pages} {076003} (\bibinfo {year} {1998})},\
  \Eprint {https://arxiv.org/abs/hep-lat/9712008} {arXiv:hep-lat/9712008}
  \BibitemShut {NoStop}%
\bibitem [{\citenamefont {Schaich}\ and\ \citenamefont
  {Loinaz}(2009)}]{Schaich:2009}%
  \BibitemOpen
  \bibfield  {author} {\bibinfo {author} {\bibfnamefont {D.}~\bibnamefont
  {Schaich}}\ and\ \bibinfo {author} {\bibfnamefont {W.}~\bibnamefont
  {Loinaz}},\ }\bibfield  {title} {\bibinfo {title} {{Improved lattice
  measurement of the critical coupling in $\phi^4_2$ theory}},\ }\href
  {https://doi.org/10.1103/PhysRevD.79.056008} {\bibfield  {journal} {\bibinfo
  {journal} {Phys. Rev. D}\ }\textbf {\bibinfo {volume} {79}},\ \bibinfo
  {pages} {056008} (\bibinfo {year} {2009})},\ \Eprint
  {https://arxiv.org/abs/0902.0045} {arXiv:0902.0045 [hep-lat]} \BibitemShut
  {NoStop}%
\bibitem [{\citenamefont {Bosetti}\ \emph {et~al.}(2015)\citenamefont
  {Bosetti}, \citenamefont {De~Palma},\ and\ \citenamefont
  {Guagnelli}}]{Bosetti:2015}%
  \BibitemOpen
  \bibfield  {author} {\bibinfo {author} {\bibfnamefont {P.}~\bibnamefont
  {Bosetti}}, \bibinfo {author} {\bibfnamefont {B.}~\bibnamefont {De~Palma}},\
  and\ \bibinfo {author} {\bibfnamefont {M.}~\bibnamefont {Guagnelli}},\
  }\bibfield  {title} {\bibinfo {title} {{Monte Carlo determination of the
  critical coupling in $\phi^4_2$ theory}},\ }\href
  {https://doi.org/10.1103/PhysRevD.92.034509} {\bibfield  {journal} {\bibinfo
  {journal} {Phys. Rev. D}\ }\textbf {\bibinfo {volume} {92}},\ \bibinfo
  {pages} {034509} (\bibinfo {year} {2015})},\ \Eprint
  {https://arxiv.org/abs/1506.08587} {arXiv:1506.08587 [hep-lat]} \BibitemShut
  {NoStop}%
\bibitem [{\citenamefont {Bronzin}\ \emph {et~al.}(2019)\citenamefont
  {Bronzin}, \citenamefont {De~Palma},\ and\ \citenamefont
  {Guagnelli}}]{Bronzin:2019}%
  \BibitemOpen
  \bibfield  {author} {\bibinfo {author} {\bibfnamefont {S.}~\bibnamefont
  {Bronzin}}, \bibinfo {author} {\bibfnamefont {B.}~\bibnamefont {De~Palma}},\
  and\ \bibinfo {author} {\bibfnamefont {M.}~\bibnamefont {Guagnelli}},\
  }\bibfield  {title} {\bibinfo {title} {{New Monte Carlo determination of the
  critical coupling in $\phi^4_2$ theory}},\ }\href
  {https://doi.org/10.1103/PhysRevD.99.034508} {\bibfield  {journal} {\bibinfo
  {journal} {Phys. Rev. D}\ }\textbf {\bibinfo {volume} {99}},\ \bibinfo
  {pages} {034508} (\bibinfo {year} {2019})},\ \Eprint
  {https://arxiv.org/abs/1807.03381} {arXiv:1807.03381 [hep-lat]} \BibitemShut
  {NoStop}%
\bibitem [{\citenamefont {Lubin}\ \emph {et~al.}(2023)\citenamefont {Lubin},
  \citenamefont {Dowson}, \citenamefont {{Dias Garcia}}, \citenamefont
  {Huchette}, \citenamefont {Legat},\ and\ \citenamefont {Vielma}}]{Lubin2023}%
  \BibitemOpen
  \bibfield  {author} {\bibinfo {author} {\bibfnamefont {M.}~\bibnamefont
  {Lubin}}, \bibinfo {author} {\bibfnamefont {O.}~\bibnamefont {Dowson}},
  \bibinfo {author} {\bibfnamefont {J.}~\bibnamefont {{Dias Garcia}}}, \bibinfo
  {author} {\bibfnamefont {J.}~\bibnamefont {Huchette}}, \bibinfo {author}
  {\bibfnamefont {B.}~\bibnamefont {Legat}},\ and\ \bibinfo {author}
  {\bibfnamefont {J.~P.}\ \bibnamefont {Vielma}},\ }\bibfield  {title}
  {\bibinfo {title} {{JuMP} 1.0: {R}ecent improvements to a modeling language
  for mathematical optimization},\ }\bibfield  {journal} {\bibinfo  {journal}
  {Mathematical Programming Computation}\ }\href
  {https://doi.org/10.1007/s12532-023-00239-3} {10.1007/s12532-023-00239-3}
  (\bibinfo {year} {2023})\BibitemShut {NoStop}%
\bibitem [{\citenamefont {ApS}(2025)}]{mosek}%
  \BibitemOpen
  \bibfield  {author} {\bibinfo {author} {\bibfnamefont {M.}~\bibnamefont
  {ApS}},\ }\href {https://docs.mosek.com/latest/juliaapi/index.html} {\emph
  {\bibinfo {title} {MOSEK Optimizer API for Julia 11.2.1}}} (\bibinfo {year}
  {2025})\BibitemShut {NoStop}%
\bibitem [{\citenamefont {Coey}\ \emph {et~al.}(2021)\citenamefont {Coey},
  \citenamefont {Kapelevich},\ and\ \citenamefont
  {Vielma}}]{coey2021solvingnaturalconicformulations}%
  \BibitemOpen
  \bibfield  {author} {\bibinfo {author} {\bibfnamefont {C.}~\bibnamefont
  {Coey}}, \bibinfo {author} {\bibfnamefont {L.}~\bibnamefont {Kapelevich}},\
  and\ \bibinfo {author} {\bibfnamefont {J.~P.}\ \bibnamefont {Vielma}},\
  }\href {https://arxiv.org/abs/2005.01136} {\bibinfo {title} {Solving natural
  conic formulations with hypatia.jl}} (\bibinfo {year} {2021}),\ \Eprint
  {https://arxiv.org/abs/2005.01136} {arXiv:2005.01136 [math.OC]} \BibitemShut
  {NoStop}%
\bibitem [{\citenamefont {Alexandrou}\ \emph {et~al.}(2023)\citenamefont
  {Alexandrou} \emph {et~al.}}]{ExtendedTwistedMassCollaborationETMC:2022sta}%
  \BibitemOpen
  \bibfield  {author} {\bibinfo {author} {\bibfnamefont {C.}~\bibnamefont
  {Alexandrou}} \emph {et~al.} (\bibinfo {collaboration} {Extended Twisted Mass
  Collaboration (ETMC)}),\ }\bibfield  {title} {\bibinfo {title} {{Probing the
  Energy-Smeared R Ratio Using Lattice QCD}},\ }\href
  {https://doi.org/10.1103/PhysRevLett.130.241901} {\bibfield  {journal}
  {\bibinfo  {journal} {Phys. Rev. Lett.}\ }\textbf {\bibinfo {volume} {130}},\
  \bibinfo {pages} {241901} (\bibinfo {year} {2023})},\ \Eprint
  {https://arxiv.org/abs/2212.08467} {arXiv:2212.08467 [hep-lat]} \BibitemShut
  {NoStop}%
\bibitem [{\citenamefont {Bergeron}\ and\ \citenamefont
  {Tremblay}(2016)}]{bergeron2016}%
  \BibitemOpen
  \bibfield  {author} {\bibinfo {author} {\bibfnamefont {D.}~\bibnamefont
  {Bergeron}}\ and\ \bibinfo {author} {\bibfnamefont {A.-M.~S.}\ \bibnamefont
  {Tremblay}},\ }\bibfield  {title} {\bibinfo {title} {Algorithms for optimized
  maximum entropy and diagnostic tools for analytic continuation},\ }\href
  {https://doi.org/10.1103/PhysRevE.94.023303} {\bibfield  {journal} {\bibinfo
  {journal} {Phys. Rev. E}\ }\textbf {\bibinfo {volume} {94}},\ \bibinfo
  {pages} {023303} (\bibinfo {year} {2016})}\BibitemShut {NoStop}%
\bibitem [{\citenamefont {Levy}\ \emph {et~al.}(2017)\citenamefont {Levy},
  \citenamefont {LeBlanc},\ and\ \citenamefont {Gull}}]{Levy_2017}%
  \BibitemOpen
  \bibfield  {author} {\bibinfo {author} {\bibfnamefont {R.}~\bibnamefont
  {Levy}}, \bibinfo {author} {\bibfnamefont {J.}~\bibnamefont {LeBlanc}},\ and\
  \bibinfo {author} {\bibfnamefont {E.}~\bibnamefont {Gull}},\ }\bibfield
  {title} {\bibinfo {title} {Implementation of the maximum entropy method for
  analytic continuation},\ }\href {https://doi.org/10.1016/j.cpc.2017.01.018}
  {\bibfield  {journal} {\bibinfo  {journal} {Computer Physics Communications}\
  }\textbf {\bibinfo {volume} {215}},\ \bibinfo {pages} {149–155} (\bibinfo
  {year} {2017})}\BibitemShut {NoStop}%
\bibitem [{\citenamefont {Kraberger}\ \emph {et~al.}(2017)\citenamefont
  {Kraberger}, \citenamefont {Triebl}, \citenamefont {Zingl},\ and\
  \citenamefont {Aichhorn}}]{PhysRevB.96.155128}%
  \BibitemOpen
  \bibfield  {author} {\bibinfo {author} {\bibfnamefont {G.~J.}\ \bibnamefont
  {Kraberger}}, \bibinfo {author} {\bibfnamefont {R.}~\bibnamefont {Triebl}},
  \bibinfo {author} {\bibfnamefont {M.}~\bibnamefont {Zingl}},\ and\ \bibinfo
  {author} {\bibfnamefont {M.}~\bibnamefont {Aichhorn}},\ }\bibfield  {title}
  {\bibinfo {title} {Maximum entropy formalism for the analytic continuation of
  matrix-valued green's functions},\ }\href
  {https://doi.org/10.1103/PhysRevB.96.155128} {\bibfield  {journal} {\bibinfo
  {journal} {Phys. Rev. B}\ }\textbf {\bibinfo {volume} {96}},\ \bibinfo
  {pages} {155128} (\bibinfo {year} {2017})}\BibitemShut {NoStop}%
\bibitem [{\citenamefont {Otsuki}\ \emph {et~al.}(2017)\citenamefont {Otsuki},
  \citenamefont {Ohzeki}, \citenamefont {Shinaoka},\ and\ \citenamefont
  {Yoshimi}}]{PhysRevE.95.061302}%
  \BibitemOpen
  \bibfield  {author} {\bibinfo {author} {\bibfnamefont {J.}~\bibnamefont
  {Otsuki}}, \bibinfo {author} {\bibfnamefont {M.}~\bibnamefont {Ohzeki}},
  \bibinfo {author} {\bibfnamefont {H.}~\bibnamefont {Shinaoka}},\ and\
  \bibinfo {author} {\bibfnamefont {K.}~\bibnamefont {Yoshimi}},\ }\bibfield
  {title} {\bibinfo {title} {Sparse modeling approach to analytical
  continuation of imaginary-time quantum monte carlo data},\ }\href
  {https://doi.org/10.1103/PhysRevE.95.061302} {\bibfield  {journal} {\bibinfo
  {journal} {Phys. Rev. E}\ }\textbf {\bibinfo {volume} {95}},\ \bibinfo
  {pages} {061302(R)} (\bibinfo {year} {2017})}\BibitemShut {NoStop}%
\bibitem [{\citenamefont {Fei}\ \emph {et~al.}(2021)\citenamefont {Fei},
  \citenamefont {Yeh},\ and\ \citenamefont {Gull}}]{fei2021}%
  \BibitemOpen
  \bibfield  {author} {\bibinfo {author} {\bibfnamefont {J.}~\bibnamefont
  {Fei}}, \bibinfo {author} {\bibfnamefont {C.-N.}\ \bibnamefont {Yeh}},\ and\
  \bibinfo {author} {\bibfnamefont {E.}~\bibnamefont {Gull}},\ }\bibfield
  {title} {\bibinfo {title} {Nevanlinna analytical continuation},\ }\href
  {https://doi.org/10.1103/PhysRevLett.126.056402} {\bibfield  {journal}
  {\bibinfo  {journal} {Phys. Rev. Lett.}\ }\textbf {\bibinfo {volume} {126}},\
  \bibinfo {pages} {056402} (\bibinfo {year} {2021})}\BibitemShut {NoStop}%
\bibitem [{\citenamefont {Kaufmann}\ and\ \citenamefont
  {Held}(2023)}]{KAUFMANN2023108519}%
  \BibitemOpen
  \bibfield  {author} {\bibinfo {author} {\bibfnamefont {J.}~\bibnamefont
  {Kaufmann}}\ and\ \bibinfo {author} {\bibfnamefont {K.}~\bibnamefont
  {Held}},\ }\bibfield  {title} {\bibinfo {title} {ana\_cont: Python package
  for analytic continuation},\ }\href
  {https://doi.org/https://doi.org/10.1016/j.cpc.2022.108519} {\bibfield
  {journal} {\bibinfo  {journal} {Computer Physics Communications}\ }\textbf
  {\bibinfo {volume} {282}},\ \bibinfo {pages} {108519} (\bibinfo {year}
  {2023})}\BibitemShut {NoStop}%
\bibitem [{\citenamefont {Mutzel}\ and\ \citenamefont
  {Tilloy}(2025)}]{Mutzel:2025veg}%
  \BibitemOpen
  \bibfield  {author} {\bibinfo {author} {\bibfnamefont {S.}~\bibnamefont
  {Mutzel}}\ and\ \bibinfo {author} {\bibfnamefont {A.}~\bibnamefont
  {Tilloy}},\ }\bibfield  {title} {\bibinfo {title} {{Extracting quantum field
  theory dynamics from an approximate ground state}},\ }\Eprint
  {https://arxiv.org/abs/2512.19594} {arXiv:2512.19594 [quant-ph]}  (\bibinfo
  {year} {2025})\BibitemShut {NoStop}%
\bibitem [{\citenamefont {Abbott}\ \emph {et~al.}(2026)\citenamefont {Abbott},
  \citenamefont {Fields}, \citenamefont {Jay}, \citenamefont {Oare},\ and\
  \citenamefont {Saccardi}}]{abbott2026causalbootstrapboundingsmeared}%
  \BibitemOpen
  \bibfield  {author} {\bibinfo {author} {\bibfnamefont {R.}~\bibnamefont
  {Abbott}}, \bibinfo {author} {\bibfnamefont {S.}~\bibnamefont {Fields}},
  \bibinfo {author} {\bibfnamefont {W.~I.}\ \bibnamefont {Jay}}, \bibinfo
  {author} {\bibfnamefont {P.}~\bibnamefont {Oare}},\ and\ \bibinfo {author}
  {\bibfnamefont {M.}~\bibnamefont {Saccardi}},\ }\href
  {https://arxiv.org/abs/2605.20509} {\bibinfo {title} {The causal bootstrap:
  Bounding smeared spectral functions from non-perturbative euclidean data}}
  (\bibinfo {year} {2026}),\ \Eprint {https://arxiv.org/abs/2605.20509}
  {arXiv:2605.20509 [hep-lat]} \BibitemShut {NoStop}%
\bibitem [{\citenamefont {Boyd}\ and\ \citenamefont
  {Vandenberghe}(2004)}]{boyd2004convex}%
  \BibitemOpen
  \bibfield  {author} {\bibinfo {author} {\bibfnamefont {S.}~\bibnamefont
  {Boyd}}\ and\ \bibinfo {author} {\bibfnamefont {L.}~\bibnamefont
  {Vandenberghe}},\ }\href {https://books.google.fr/books?id=IUZdAAAAQBAJ}
  {\emph {\bibinfo {title} {Convex Optimization}}}\ (\bibinfo  {publisher}
  {Cambridge University Press},\ \bibinfo {year} {2004})\BibitemShut {NoStop}%
\bibitem [{\citenamefont {Nocedal}\ and\ \citenamefont
  {Wright}(2006)}]{NocedalWright}%
  \BibitemOpen
  \bibfield  {author} {\bibinfo {author} {\bibfnamefont {J.}~\bibnamefont
  {Nocedal}}\ and\ \bibinfo {author} {\bibfnamefont {S.~J.}\ \bibnamefont
  {Wright}},\ }\href@noop {} {\emph {\bibinfo {title} {Numerical
  optimization}}},\ \bibinfo {edition} {2nd}\ ed.,\ Springer series in
  operations research and financial engineering\ (\bibinfo  {publisher}
  {Springer},\ \bibinfo {year} {2006})\BibitemShut {NoStop}%
\bibitem [{\citenamefont {Marčenko}\ and\ \citenamefont
  {Pastur}(1967)}]{Marcenko_1967}%
  \BibitemOpen
  \bibfield  {author} {\bibinfo {author} {\bibfnamefont {V.~A.}\ \bibnamefont
  {Marčenko}}\ and\ \bibinfo {author} {\bibfnamefont {L.~A.}\ \bibnamefont
  {Pastur}},\ }\bibfield  {title} {\bibinfo {title} {Distribution of
  eigenvalues for some sets of random matrices},\ }\href
  {https://doi.org/10.1070/SM1967v001n04ABEH001994} {\bibfield  {journal}
  {\bibinfo  {journal} {Mathematics of the USSR-Sbornik}\ }\textbf {\bibinfo
  {volume} {1}},\ \bibinfo {pages} {457} (\bibinfo {year} {1967})}\BibitemShut
  {NoStop}%
\bibitem [{\citenamefont {Yoon}\ \emph {et~al.}(2013)\citenamefont {Yoon},
  \citenamefont {Jang}, \citenamefont {Jung},\ and\ \citenamefont
  {Lee}}]{Yoon:2011wdw}%
  \BibitemOpen
  \bibfield  {author} {\bibinfo {author} {\bibfnamefont {B.}~\bibnamefont
  {Yoon}}, \bibinfo {author} {\bibfnamefont {Y.-C.}\ \bibnamefont {Jang}},
  \bibinfo {author} {\bibfnamefont {C.}~\bibnamefont {Jung}},\ and\ \bibinfo
  {author} {\bibfnamefont {W.}~\bibnamefont {Lee}},\ }\bibfield  {title}
  {\bibinfo {title} {{Covariance fitting of highly correlated data in lattice
  QCD}},\ }\href {https://doi.org/10.3938/jkps.63.145} {\bibfield  {journal}
  {\bibinfo  {journal} {J. Korean Phys. Soc.}\ }\textbf {\bibinfo {volume}
  {63}},\ \bibinfo {pages} {145} (\bibinfo {year} {2013})},\ \Eprint
  {https://arxiv.org/abs/1101.2248} {arXiv:1101.2248 [hep-lat]} \BibitemShut
  {NoStop}%
\bibitem [{\citenamefont {Bruno}\ and\ \citenamefont
  {Sommer}(2023)}]{Bruno:2022mfy}%
  \BibitemOpen
  \bibfield  {author} {\bibinfo {author} {\bibfnamefont {M.}~\bibnamefont
  {Bruno}}\ and\ \bibinfo {author} {\bibfnamefont {R.}~\bibnamefont {Sommer}},\
  }\bibfield  {title} {\bibinfo {title} {{On fits to correlated and
  auto-correlated data}},\ }\href {https://doi.org/10.1016/j.cpc.2022.108643}
  {\bibfield  {journal} {\bibinfo  {journal} {Comput. Phys. Commun.}\ }\textbf
  {\bibinfo {volume} {285}},\ \bibinfo {pages} {108643} (\bibinfo {year}
  {2023})},\ \Eprint {https://arxiv.org/abs/2209.14188} {arXiv:2209.14188
  [hep-lat]} \BibitemShut {NoStop}%
\end{thebibliography}%
